\documentclass[12pt,preprint]{elsarticle}
\usepackage{amsmath}
\usepackage{graphics,graphicx}
\usepackage{morefloats}
\usepackage{subfig}
\usepackage{multirow}
\usepackage{array}
\usepackage{rotating}
\usepackage[T1]{fontenc}
\usepackage{textcomp}
\usepackage[hyphens]{url}
\usepackage{hyperref}
\newcommand{\qvec}[1]{\boldsymbol{#1}}
\newcommand*\icarus{Icarus~}
\newcommand*\jgr{J Geophys Res~}
\newcommand*\grl{Geophys Res Lett}
\newcommand*\planss{Planetary Space Sci}
\newcommand*\jqsrt{J Quant Spec Rad Trans}
\journal{JQRST}
\begin{document}
\begin{frontmatter}
%

%%% Title %%%
\title{Linearized Flux Evolution (LiFE): A Technique for Rapidly Adapting Fluxes from Full-Physics Radiative Transfer Models}
%%%

%%% Authors %%%
\author{Tyler D. Robinson$^{a,b}$}
\address{Department of Physics and Astronomy, Northern Arizona University, Flagstaff, AZ 86011, USA; tyler.robinson@nau.edu}

%\and

\author{David Crisp$^{a}$}
\address{Jet Propulsion Laboratory, California Institute of Technology, Pasadena, CA 91109, USA}

\address[label1]{NASA Astrobiology Institute's Virtual Planetary Laboratory}
%\address[label2]{Sagan Fellow}
\address[label2]{University of California, Santa Cruz, Other Worlds Laboratory}
%%%

%
\begin{abstract} 
Solar and thermal radiation are critical aspects of planetary climate, with gradients in radiative 
energy fluxes driving heating and cooling.  Climate models require that radiative transfer tools 
be versatile, computationally efficient, and accurate.  Here, we describe a technique that 
uses an accurate full-physics radiative transfer model to generate a set of atmospheric radiative 
quantities which can be used to linearly adapt radiative flux profiles to changes in the 
atmospheric and surface state---the Linearized Flux Evolution (LiFE) approach.  These 
radiative quantities describe how each model layer in a plane-parallel atmosphere reflects 
and transmits light, as well as how the layer generates diffuse radiation by thermal emission 
and by scattering light from the direct solar beam.  By computing derivatives of these layer 
radiative properties with respect to dynamic elements of the atmospheric state, we can 
then efficiently adapt the flux profiles computed by the full-physics model to new atmospheric 
states.  We { validate the LiFE approach}, and then apply this approach to Mars, 
Earth, and Venus, demonstrating the information contained in the layer radiative properties 
and their derivatives, as well as how the LiFE approach can be used to determine the thermal 
structure of radiative and radiative-convective equilibrium states in one-dimensional atmospheric 
models.
\end{abstract}
%

%
%\begin{keyword}
%

%
%\end{keyword}
%

%
\end{frontmatter}
%

%%%
\section{Introduction}
%%%

Plane-parallel, horizontally homogenous radiative transfer calculations are the standard 
approach to exploring solar and thermal radiation fields in Earth and planetary atmospheres.  
In the context of planetary climate models, which derive an equilibrium or quasi-equilibrium 
surface-atmosphere state from a specified set of initial conditions,  it is the role of a radiative 
transfer model to compute the state-dependent vertical net radiative energy flux.  Gradients in the 
net radiative flux contribute to atmospheric heating or cooling, which the climate model uses to 
update the surface and atmospheric state \citep{manabe&strickler1964,manabe&wetherald1967,
ramanthan&coakley1978,schlesinger&mitchell1987}.  Since the net radiative flux is a function of 
these state properties, the radiative transfer model must then re-compute the radiative fluxes 
for any new atmospheric state.  This modeling approach demands that the radiative transfer 
model be both accurate and computationally efficient 
\citep[e.g.,][]{pollack&ackerman1983,lacisetal1990}.

While a number of techniques have been developed to accurately compute radiative energy 
fluxes in a realistic, vertically-inhomogeneous planetary atmosphere where gases and airborne 
particles contribute to the absorption, emission, and multiple scattering of solar or thermal 
radiation \citep{hansen&travis1974,stamnesetal1988,spurretal2001}, 
these approaches are typically too computationally expensive to be used within a general climate 
model, which requires fluxes to be computed and integrated over a large range of wavelengths 
(or frequencies).  Thus, climate models tend to rely on parametrized radiative transfer tools 
\citep{goodyetal1989,lacis&oinas1991,kiehl&briegleb1991,brieglebetal1992,mlaweretal1997,
bullock&grinspoon2001,iaconoetal2008,wordsworthetal2010}.  However, the tuning and 
parameterizations used in these tools can lead to errors and biases in predictions when the 
model is used to study conditions for which it was not designed.

Here, we describe a technique for efficiently adapting radiative fluxes, computed with  a 
full-physics radiative transfer model, to changes in the atmospheric state---the LiFE approach.  
{ This technique combines linear ``flux Jacobians'' with an approximate method motivated by 
two-stream adding methods \citep{briegleb1992,shettle&weinman1970,crisp1986} to adapt the radiation 
field to changes in the atmospheric and surface thermal structure and optical properties as they 
evolve.}  Thus, the LiFE approach is applicable to a wide range of problems in planetary climate.

In what follows, we begin with a description of the LiFE approach 
(Section~\ref{sec:modeldescription}), including a discussion of how we compute the requisite layer 
radiative properties and the flux Jacobians associated with each property (Sections~\ref{sec:layprops} 
and \ref{sec:propderiv}, respectively).  { We then present a validation of the LiFE approach in 
Section~\ref{sec:validation}.}  Next we demonstrate a sequence of example applications of the 
LiFE approach (Section~\ref{sec:examples}), which focus on Mars (Section~\ref{sec:mars}), 
Earth (Section~\ref{sec:earth}), and Venus (Section~\ref{sec:venus}).  { For a simple test case, 
Section~\ref{sec:comparison} shows a comparison of a LiFE-derived atmospheric thermal profile to an 
atmospheric state computed using a widely-adopted, one-dimensional radiative-convective planetary 
climate model.}  Note that the Appendix contains an intuitive example calculation using the 
{ approximate} two-stream flux adding technique, and, for completeness, clear derivations of the 
flux adding relationships adopted for this application.
 
%%%
\section{Description of LiFE Approach}
\label{sec:modeldescription}
%%%

The Linearized Flux Evolution (LiFE) approach pairs a full-physics radiative transfer model 
with { flux Jacobians and an approximate method motivated by two-stream flux adding techniques 
\citep{briegleb1992,shettle&weinman1970,crisp1986}} to rapidly adapt radiative flux profiles in 
plane-parallel planetary atmospheres to changes in atmospheric state.  Given an initial atmospheric 
and surface state, the full-physics model is used to compute the upwelling and downwelling radiative 
energy fluxes at the boundaries between model atmospheric layers.  Additionally, for each atmospheric 
layer, the full-physics model computes a set of frequency-dependent layer radiative properties and their 
flux Jacobians, which are used to determine the first derivative of the upward and downward solar and 
thermal fluxes at each layer boundary with respect to changes in the atmospheric or surface thermal 
structure or optical properties.  

{ In the LiFE approach, we decompose a vertically inhomogeneous, non-isothermal atmosphere into a 
series of discrete interacting layers.  Each layer is characterized by an effective flux reflectivity, transmissivity, 
and upwelling and downwelling ``source terms'' (collectively referred to as ``layer radiative properties'').  While 
adding approaches are most correctly applied to radiances \citep[][Chapter VII]{chandrasekhar1960}, we 
specifically construct these layer radiative properties to reproduce the level-dependent fluxes from a 
full-physics model when used within our flux adding method.  The accuracy of the computed Jacobians for 
the layer radiative properties (and, thus, the accuracy of the flux adding approach) decays as the atmosphere 
and surface evolve from their initial state, eventually requiring a recalculation of the flux field and layer 
Jacobians using the full-physics model.}

{ Unlike analytic two-stream adding methods \citep[e.g.,][]{shettle&weinman1970}, the layer radiative 
properties used here are not derived from analytic two-stream 
solutions to the equation of transfer.  Instead, these properties are {\it defined} such that the two-stream adding 
method exactly reproduces the flux distributions generated by the full-physics model for a specified 
atmospheric state. Similarly, the reflectivity, transmissivity, and source term Jacobians are defined to 
approximate the linear response of the fluxes in the full-physics model to small perturbations in the layer 
optical properties (e.g., differential optical depth, single scattering albedo, scattering phase function) 
associated with changes in temperature, or trace gas and aerosol profiles.}

%The frequency-dependent radiative properties for each model layer --- its reflectivity, transmissivity, 
%and source terms --- are the necessary inputs to a flux adding technique, which solve for the upwelling 
%and downwelling diffuse radiative fluxes.  The derivatives of these properties form Jacobians which 
%(a matrix of derivatives for each layer radiative property with dimensions equal to the number of spectral bins by the number of atmospheric layers) 
%are used to adapt the layer radiative properties to changes in the atmospheric state using a linear 
%(first-order) Taylor expansion.  The Jacobians can be thought of as providing an estimate of how 
%the radiative flux profile will respond to a change in the atmospheric and surface state vector, 
%$\qvec{x}$ (which is generally a function of altitude and frequency).  Thus, if $F_{\nu}(\qvec{x})$ is the 
%frequency-dependent radiative flux profile for the initial atmospheric state, as computed by the full-physics 
%model, and the atmospheric state evolves to $\qvec{x} + \delta\qvec{x}$, then we can use the Jacobians 
%and the flux adding technique to compute $\partial F_{\nu}/\partial \qvec{x}$, and then adjust the original 
%flux profile according to
%\begin{equation}
%F_{\nu}(\qvec{x}+ \delta\qvec{x}) \approx F_{\nu}(\qvec{x}) + 
%    \frac{\partial F_{\nu}}{\partial \qvec{x}}\delta\qvec{x} \ .
%\label{eqn:example}
%\end{equation}

{ In general, bolometric or band-integrated radiances and fluxes have a non-linear dependence on the 
atmospheric thermal structure and optical properties. However, by applying a linear flux correction over 
monochromatic (or near-monochromatic) spectral intervals (i.e., of order 1--10~cm$^{-1}$), we aim to minimize 
sensitivity within the LiFE approach to the non-linear effects that arise in wavelength-integrated quantities.}  Our 
linear flux correction approach was motivated by atmospheric retrieval algorithms.  Within a retrieval framework 
\citep{rodgers2000}, a synthetic spectrum and associated linear radiance Jacobians are generated for an assumed 
atmospheric and surface state.  These results are then used within the context of a non-linear least-squares fitting 
algorithm to produce a better fit to an observed spectrum.  Here, like in the LiFE approach, the non-linear relationship 
between radiances (or fluxes) and optical properties is approximated as quasi-linear for small perturbations 
to the atmospheric and surface state.

In the following sub-sections, we first present the flux adding technique, which will introduce 
the layer radiative properties relevant to LiFE.  We then discuss how we use a full-physics 
radiative transfer model to compute the layer radiative properties.  Finally, we discuss how 
we compute the Jacobians and how these are used to rapidly adapt radiative flux profiles as an 
atmospheric state evolves from its initial conditions toward an equilibrium state.  While these 
discussions typically apply to a narrow spectral range within a spectrally resolved 
model, an intuitive example of the LiFE approach which uses gray radiative transfer is 
provided in the Appendix to help the reader better understand the technique.

%%%
\subsection{Flux Adding}
\label{sec:flxadding}
%%%

The flux adding technique relates the diffuse upwelling and downwelling radiative energy fluxes, 
defined at the boundaries between a collection of homogenous atmospheric layers, to the radiative 
properties of each of these layers \citep{stephens1976,crisp1986,harshvardhanking1993}. By stacking 
homogenous layers to create inhomogeneous layers, we can recursively relate the radiative fluxes to 
the properties of successively thicker atmospheric layers, eventually determining the fluxes at all 
required levels throughout the atmosphere.  This process is analogous to previous two-stream 
adding techniques \citep{briegleb1992,shettle&weinman1970,crisp1986}.

Each homogenous layer in a model atmosphere is assigned a reflectivity and transmissivity, 
which describe how the layer reflects and transmits diffuse incident flux.  Additionally, each layer 
can add radiation to the diffuse flux field, either by scattering light from a direct solar 
beam, or by emitting thermal radiation.  A layer's ability to add diffuse flux is described 
by layer source terms, which are not necessarily identical for contributions to the 
upwelling and downwelling diffuse flux fields.  The frequency-dependent diffuse flux 
reflectivity, transmissivity, and upwelling and downwelling source terms for layer 
$j$ will be labeled $r_{j}$, $t_{j}$, $s^{+}_{j}$, and $s^{-}_{j}$, respectively.  The diffuse 
flux emerging from layer $j$, $F^{+}_{j-1}$ and $F^{-}_{j}$, is related to the incident diffuse 
flux, $F^{-}_{j-1}$ and $F^{+}_{j}$, through
\begin{equation}
F^{+}_{j-1} = r_{j}F^{-}_{j-1} + t_{j}F^{+}_{j} + s^{+}_{j} \ ,
\label{eqn:adding_j_up}
\end{equation}
\begin{equation}
F^{-}_{j} = t_{j}F^{-}_{j-1} + r_{j}F^{+}_{j} + s^{-}_{j} \ .
\label{eqn:adding_j_dn}
\end{equation}
Figure~\ref{fig:adding_single} shows a schematic of these layer properties and fluxes.

As is derived in the Appendix, two homogenous layers with different radiative properties can 
be combined to yield an inhomogeneous layer.  The radiative properties of subsequently thicker 
inhomogeneous layers can be obtained by adding homogenous layers to either the bottom or top 
of an inhomogeneous layer.  These downward and upward adding procedures produce a 
recursive set of relations that describe how the inhomogeneous layers reflect and 
generate diffuse flux.

When adding downwards, homogenous layer $j$ is added to the base of an inhomogeneous 
layer that extends from the top of the atmosphere ($j=0$) to layer $j-1$.  We define the 
inhomogeneous layer reflectivity and source term for adding downward, $R^{-}_{0,j-1}$ and 
$S^{-}_{0,j-1}$, respectively, and have 
\begin{equation}
F^{-}_{j-1} = R^{-}_{0,j-1}F^{+}_{j-1} + S^{-}_{0,j-1} \ ,
\label{eqn:inhomo_layer_down_j-1}
\end{equation}
which are generally frequency-dependent, and are schematically shown in 
Figure~\ref{fig:adding_down}.  The reflectivity and source term for the newly-extended 
inhomogeneous layer are then given by (see Appendix)
\begin{equation}
R^{-}_{0,j} = r_{j} + \frac{t_{j}^{2}R^{-}_{0,j-1}}{1 - r_{j}R^{-}_{0,j-1}} \ ,
\label{eqn:inhomo_refl_down}
\end{equation}
and
\begin{equation}
S^{-}_{0,j} = s^{-}_{j} + \frac{t_{j}\left( S^{-}_{0,j-1} + s^{+}_{j}R^{-}_{0,j-1} \right)}{1 - r_{j}R^{-}_{0,j-1}} \ .
\label{eqn:inhomo_source_down}
\end{equation}

Similarly, in upwards adding, homogenous layer $j$ is added to the top of an 
inhomogeneous layer extending from the base of the atmosphere $j=N-1$ to the 
bottom of layer $j$.  For this scenario, we have
\begin{equation}
F_{j}^{+} = R_{j+1,N}^{+}F_{j}^{-} + S_{j+1,N}^{+}
\label{eqn:inhomo_layer_up}
\end{equation}
where we have defined the inhomogeneous layer reflectivity and source term for adding 
upward as $R_{j+1,N}^{+}$ and $S_{j+1,N}^{+}$, respectively.  Figure~\ref{fig:adding_up} shows 
a schematic of the process of adding a homogenous layer to the base of an inhomogeneous layer.
The recursive relations for upward adding are then given by (see Appendix) 
\begin{equation}
R_{j,N}^{+} = r_{j} + \frac{t_{j}^{2}R^{+}_{j+1,N}}{1 - r_{j}R^{+}_{j+1,N}} \ ,
\label{eqn:inhomo_refl_up}
\end{equation}
and
\begin{equation}
S_{j,N}^{+} = s^{+}_{j} + \frac{t_{j}\left( S^{+}_{j+1,N} + s^{-}_{j}R^{+}_{j+1,N} \right)}{1 - r_{j}R^{+}_{j+1,N}} \ .
\label{eqn:inhomo_source_up}
\end{equation}

Note that a convenient expression can be found by inserting the equivalent expression to 
Equation~\ref{eqn:inhomo_layer_down_j-1} for an inhomogeneous layer extending down to layer $j$ 
into Equation~\ref{eqn:inhomo_layer_up}.  After simplifying, this gives us
\begin{equation}
F_{j}^{+} = \frac{ S^{+}_{j+1,N} + R_{j+1,N}^{+}S^{-}_{1,j} }{1 - R_{j+1,N}^{+}R_{1,j}^{-}} \ .
\label{eqn:upflx_adding}
\end{equation}
This relation provides the upwelling flux at each model level in terms of quantities that can be 
recursively determined from only the homogenous layer radiative properties.  { The equivalent 
expression for downwelling flux is determined via a similar insertion, and is given by}
\begin{equation}
F_{j-1}^{-} = \frac{ S^{-}_{0,j-1} + R_{0,j-1}^{-}S^{+}_{j,N} }{1 - R_{0,j-1}^{-}R_{j,N}^{+}} \ .
\label{eqn:dnflx_adding}
\end{equation}

The relations described above allow us to compute the upwelling and downwelling fluxes at each model 
level (i.e., at the boundaries between homogenous atmospheric layers) when given the radiative 
properties for each homogenous model atmospheric layer.  Top- and bottom-of-atmosphere boundary 
conditions must also be provided.  Here, we adopt the same boundary conditions that are used by the 
full-physics model: (1) the reflectivity of ``space" is zero (i.e., $r_{0} = R^{-}_{0,0} = 0$), (2) the 
top-of-atmosphere downwelling flux is known ($F^{-}_{0}=s^{-}_{0}=S^{-}_{0,0}$), (3) a 
frequency-dependent surface albedo is known ($A_{\nu}$, with $r_{N}=R^{+}_{N,N}=A_{\nu}$),  and 
(4) { a bottom-of-atmosphere upwelling flux is set [e.g.,  
$s^{+}_{N}=S^{+}_{N,N} = A_{\nu}S^{-}_{0,N} + (1-A_{\nu})\pi B_{\nu}(T_{s})$, 
where the first term on the right-hand side represents reflected downwelling radiation and the second 
term represents surface emission via a Planck function at the surface temperature, $T_{s}$, multiplied 
by the surface emissivity, which is $1-A_{\nu}$ by Kirchoff's Law]}.  The steps required in computing $F^{+}_{j}$ 
and $F^{-}_{j}$ within a particular spectral interval and for all values of $j$ are then:
\begin{enumerate}
  \item use the top-of-atmosphere boundary conditions, $R^{-}_{0,0} = 0$ and 
        $S^{-}_{0,0} = F^{-}_{0}$, to recursively add layers downward, using 
        Equations~\ref{eqn:inhomo_refl_down} and \ref{eqn:inhomo_source_down} to solve 
        for $R^{-}_{0,j}$ and $S^{-}_{0,j}$ for all values of $j$;
  \item use the bottom-of-atmosphere boundary conditions, $R^{+}_{N,N}=A_{\lambda}$,  
        and $S^{+}_{N,N} = A_{\lambda}S^{-}_{0,N}$ or 
        $S^{+}_{N,N} = (1-A_{\lambda})\pi B_{\lambda}(T_{s})$, to recursively add layers 
        upward, using Equations~\ref{eqn:inhomo_refl_up} and \ref{eqn:inhomo_source_up} 
        to solve for $R^{+}_{j,N}$ and $S^{+}_{j,N}$ for all values of $j$;
  \item use the computed sets of inhomogeneous layer radiative properties to compute 
        the upwelling flux at each model level using Equation~\ref{eqn:upflx_adding}; 
  \item use the upwelling fluxes and the inhomogeneous layer radiative properties to 
        compute the downwelling fluxes at each model level using 
        Equation~\ref{eqn:inhomo_layer_down_j-1}.
\end{enumerate}

%%%
\subsection{Determining Homogenous Layer Radiative Properties}
\label{sec:layprops}
%%%

A full-physics radiative transfer model (e.g., a multiple scattering, spectrum resolving [line-by-line], 
multi-stream radiative transfer model) is used to compile the optical properties of the surface and 
atmosphere at the beginning of each experiment. It then derives the angle-dependent radiances and 
fluxes throughout the atmospheric column.  The full-physics model is then used to obtain the layer 
reflectivity, transmissivity, and source terms.  These properties and fluxes are defined on a spectral 
grid that is fine enough to resolve all of the frequency-dependent changes in the atmospheric or 
surface optical properties.

Given an the atmospheric state, $\qvec{x}$, the full-physics model solves the one-dimensional, 
plane-parallel radiative transfer equation at frequency $\nu$ in an inhomogeneous scattering, 
absorbing, and emitting atmosphere, which is given by
\begin{equation}
\mu \frac{dI_{\nu}}{d\tau_{\nu}} = 
          I_{\nu}\left(\tau_{\nu},\mu,\phi,\qvec{x}\right) 
          - S_{\nu}\left(\tau_{\nu},\mu,\phi,\qvec{x}\right) \ ,
\label{eqn:rte}
\end{equation}
where $I_{\nu}$ is the spectral radiance, 
$\tau_{\nu}=\tau_{\nu}\left(\qvec{x}\right)$ is the frequency-dependent extinction 
optical depth (which increases towards higher pressures), $\mu$ is the cosine of the zenith 
angle, and $\phi$ is the azimuth angle.  The source function, $S_{\nu}$, is given by 
\scriptsize
\begin{equation}
\begin{array}{ll}
S_{\nu}\left(\tau_{\nu},\mu,\phi,\qvec{x}\right) = & 
        \omega_{\nu} F^{\odot}_{\nu} e^{-\tau_{\nu}/\mu_{\odot}} \cdot
        P_{\nu}\left(\tau_{\nu},\mu,\phi,-\mu_{\odot},\phi_{\odot},\qvec{x}\right)/4\pi  \\ & 
        + \left(1 - \omega_{\nu}\right) B_{\nu}\left(T\left(\tau_{\nu}\right)\right) \\ & 
        + \omega_{\nu} \int_{0}^{2\pi} d\phi^{\prime} \int_{-1}^{1} d\mu^{\prime} \cdot 
         I_{\nu}\left(\tau_{\nu},\mu^{\prime},\phi^{\prime},\qvec{x}\right) 
         P_{\nu}\left(\tau_{\nu},\mu,\phi,\mu^{\prime},\phi^{\prime},\qvec{x}\right)/4\pi \ ,
\end{array}
\label{eqn:defn_sourcefunc}
\end{equation}
\normalsize
where $ \omega_{\nu} = \omega_{\nu}\left( \tau_{\nu} \right) $ is the 
frequency-dependent single scattering albedo, $F^{\odot}_{\nu}$ is the 
top-of-atmosphere solar irradiance, $\mu_{\odot}$ is the solar zenith angle, $\phi_{\odot}$ 
is the solar azimuth angle, $P_{\nu}$ is the scattering phase function, $B_{\nu}$ 
is the Planck function, and $T\left(\tau_{\nu}\right)$ is the atmospheric temperature 
profile.  A pair of boundary conditions are needed to solve the radiative transfer equation, 
and these typically specify the downwelling radiation field at the top of the model 
atmosphere (i.e., $I_{\nu}\left(\tau_{\nu}=0,\mu,\phi\right)$ for $\mu < 0$) and the 
upwelling radiation field at the bottom of the atmosphere (i.e., 
$I_{\nu}\left(\tau_{\nu}=\tau_{\nu}^{*},\mu,\phi\right)$ for $\mu > 0$, where 
$\tau_{\nu}^{*}$ is the total extinction optical depth of the atmosphere at 
frequency $\nu$).

{ In principal, the layer radiative properties can be defined explicitly, using a stream-by-stream 
definition, like those used in adding/doubling methods \citep{vandehulst1963,twomeyetal1966,
granthunt1968,hansen1969a,wiscombe1976}. However, this would be far too computationally 
expensive for our application, as values would need to be computed over multiple streams at 
a large number of points throughout the solar and infrared spectral ranges.  Thus, we adopt an 
approximate method described in the following paragraphs.}

At each spectral grid point, we determine the { diffuse} transmissivity and reflectivity of 
each atmospheric layer by illuminating the layer from above with a diffuse flux, 
$F_{0}$ (typically taken as 1 W m$^{-2}$ $\mu$m$^{-1}$).  We solve the 
radiative transfer equation for the homogenous layer 
\citep[see ][]{stamnes&swanson1981} with $B_{\nu}=0$ and 
$F^{\odot}_{\nu}=0$, subject to the boundary conditions at the top of the 
layer ($\tau_{\nu}=0$) and the bottom of the layer ($\tau_{\nu}=\Delta \tau_{\nu,j}$, 
where $\Delta \tau_{\nu,j}$ is the total extinction optical depth of the layer) that
\begin{equation}
\begin{array}{lr}
I_{\nu}\left(\tau_{\nu},\mu,\phi,\qvec{x} \right) \rvert_{\tau_{\nu}=0} = F_{0}/\pi & \forall \mu < 0 \ \ \\ 
I_{\nu}\left(\tau_{\nu},\mu,\phi,\qvec{x} \right) \rvert_{\tau_{\nu}=\Delta \tau_{\nu,j}} = 0 & \forall \mu > 0 \ .
\end{array}
\end{equation}
The layer transmissivity and reflectivity are then {defined by} 
\begin{equation}
t_{j} = - \frac{2\pi}{F_{0}}\int_{-1}^{0} d\mu^{\prime} 
\cdot \mu^{\prime}I_{\nu}\left(\mu^{\prime},\qvec{x} \right) \rvert_{\tau_{\nu}=\Delta \tau_{\nu,j}} \ ,
\label{eqn:defn_tj}
\end{equation}
\begin{equation}
r_{j} = \frac{2\pi}{F_{0}} \int_{0}^{1} d\mu^{\prime} 
\cdot \mu^{\prime}I_{\nu}\left(\mu^{\prime},\qvec{x} \right) \rvert_{\tau_{\nu}=0} \ .
\label{eqn:defn_rj}
\end{equation}
{ Note that the azimuthal component of the expressions above has been integrated directly as 
the illumination is diffuse.}  Also, as is discussed in our applications below, we use a discrete ordinate approach 
\citep[{\tt DISORT},][]{stamnesetal1988} in our full-physics tool, so that 
Equations~\ref{eqn:defn_tj} and \ref{eqn:defn_rj} are, in practice, implemented using Gaussian 
quadrature to integrate over the zenith angle.

The integral in Equation~\ref{eqn:defn_tj} is the downwelling flux coming from the bottom of the 
diffusely-illuminated layer, and the integral in Equation~\ref{eqn:defn_rj} is the upwelling flux 
at the top of the layer.  Thus, the layer transmissivity is the fraction of the diffuse flux that is 
transmitted through the homogeneous layer, and the layer reflectivity is the fraction of the diffuse 
flux that is reflected backwards.  Whatever portion of $F_{0}$ that is not transmitted or reflected 
must be absorbed by the layer, so we also define a layer absorptivity as
\begin{equation}
a_{j} = 1 - t_{j} - r_{j} \ .
\end{equation}

To determine the frequency-dependent layer source terms, $s^{+/-}_{j}$, we use the 
direct and diffuse radiative flux profiles with the full-physics model.  Integrating the radiances 
from the full-physics solution of the radiative transfer equation over the upper and lower 
hemispheres of solid angle yields the upwelling and downwelling spectral radiative energy fluxes 
at each level in the model atmosphere, or
\begin{equation}
F^{\uparrow}_{\nu}\left(\tau_{\nu},\qvec{x}\right) = \int_{0}^{2\pi} d\phi^{\prime} 
             \int_{0}^{1} d\mu^{\prime} \cdot \mu^{\prime}I\left(\tau_{\nu},\mu^{\prime},\phi^{\prime},\qvec{x}\right) \ ,
\end{equation}
\begin{equation}
F^{\downarrow}_{\nu}\left(\tau_{\nu},\qvec{x}\right) = -\int_{0}^{2\pi} d\phi^{\prime} 
             \int_{-1}^{0} d\mu^{\prime} \cdot \mu^{\prime}I\left(\tau_{\nu},\mu^{\prime},\phi^{\prime},\qvec{x}\right) \ .
\end{equation}
We split these upwelling and downwelling fluxes into three components: a direct downwelling 
flux, $F^{-,di}$ (only applicable to solar sources), a diffuse upwelling flux, $F^{+}$, and a 
diffuse downwelling flux, $F^{-}$, which give
\begin{equation}
F^{\uparrow} = F^{+} \ ,
\end{equation}
\begin{equation}
F^{\downarrow} = F^{-,di} + F^{-} \ .
\end{equation}
Note that these terms are frequency-dependent, but we have dropped the sub-script $\nu$ for cleaner 
presentation and to be consistent with the discussion in the flux adding discussion.  

Given the upwelling and downwelling diffuse flux profiles and the layer radiative properties, 
the homogenous layer source terms can be determined from Equations~\ref{eqn:adding_j_up} and 
\ref{eqn:adding_j_dn}.  Rearranging these two expressions yields
\begin{equation}
s^{+}_{j} = F^{+}_{j-1} - r_{j}F^{-}_{j-1} + t_{j}F^{+}_{j} \ ,
\label{eqn:layer_sup}
\end{equation}
\begin{equation}
s^{-}_{j} = F^{-}_{j} - t_{j}F^{-}_{j-1} + r_{j}F^{+}_{j} \ ,
\label{eqn:layer_sdn}
\end{equation}
{ where we note that radiation scattered from the direct beam enters the diffuse field and, 
as a result, appears in the layer source terms.}  {The layer source terms, $s^{+}_{j}$ and $s^{-}_{j}$, 
when used with the layer transmissivities and reflectivities within our flux adding approach, are 
designed to exactly reproduce the level-dependent diffuse upwelling and downwelling radiative 
fluxes that were determined by the full-physics model.  Thus, while adding techniques are most 
correctly applied to  radiance streams, we collect any errors in our flux adding approach into the 
layer source terms.}

The layer source terms will depend on whether or not solar or thermal sources are included in 
the full-physics calculation.  When considering thermal sources, the layer source terms will be 
strong functions of temperature.  As temperature is typically a rapidly changing part of the atmospheric 
state (e.g., as a climate model marches forward in time), the thermal layer source terms can vary 
strongly in time as atmospheric temperatures evolve.  To increase the versatility of the thermal layer 
source terms and to minimize sensitivity to atmospheric temperature variations, we subtract a term 
that resembles the layer Planck emission from the layer source terms to produce a adjusted thermal 
source.  

{ In the weak absorption limit}, the Planck-like contribution to the layer source terms will resemble 
$a_{j} \pi \left( B_{\nu}(T_j) + B_{\nu}(T_{j-1}) \right)/2$, where { $a_{j}$ behaves like a layer 
emissivity,} $T_{j}$ is the temperature at level $j$ in the atmosphere, and the final term in this 
expression behaves like an average Planck function for layer $j$.  However, in the strong 
absorption limit, where the layer transmissivity goes to zero and absorptivity to unity (i.e., $t_{j} \rightarrow 0$ and 
$a_{j} \rightarrow 1$), the upwelling and downwelling fluxes at level $j$ approach the layer 
source terms $s^{+}_{j+1}$ and $s^{-}_{j}$, respectively.  In these conditions, we expect 
that the radiation diffusion expression applies, where $F_{net} \propto dB_{\nu}(T)/d\tau$.  To 
ensure that our definition of the Planck-like contribution to the layer sources yields the 
diffusion limit when the layer absorptivity is large, we adopt a ``linear in $\tau$'' form 
\citep[e.g.][their Equation~9]{cloughetal1992} of our layer thermal sources, with
\footnotesize
\begin{equation}
%\begin{array}{ll}
%\tilde{s}^{+}_{th,j} = & s^{+}_{j} - \pi a_{j} \Big[ \frac{1}{2} \left( 1 - a_{j} \right) B_{\nu}(T_{j}) 
%                                                                     + \frac{1}{2} \left( 1 + a_{j} \right) B_{\nu}(T_{j-1}) \\ &
%                                                                     - \Big(B_{\nu}(T_{j}) - B_{\nu}(T_{j-1})\Big) \left(1 - a_{j} + \frac{a_{j}}{\ln\left(1-a_{j}\right)}\right) \Big] \ ,
%\tilde{s}^{+}_{th,j} = & s^{+}_{j} - \pi \Big[ a_{j} B_{\nu}(T_{j}) -  \Big(B_{\nu}(T_{j}) - B_{\nu}(T_{j-1})\Big) \left(1 - a_{j}\right) \\ &
%                                                                + a_{j} \frac{B_{\nu}(T_{j}) - B_{\nu}(T_{j-1}}{\ln\left(1-a_{j}\right)} \Big] \ ,
\tilde{s}^{+}_{th,j} =  s^{+}_{j} - \pi \Big[ a_{j} B_{\nu}(T_{j}) -  \Big(B_{\nu}(T_{j}) - B_{\nu}(T_{j-1})\Big) \left(1 - a_{j}\right) 
                                                                - a_{j} \frac{B_{\nu}(T_{j}) - B_{\nu}(T_{j-1})}{\ln\left(1-a_{j}\right)} \Big] \ ,
%\end{array}
\label{eqn:source_scaled_thup}
\end{equation}
\begin{equation}
%\begin{array}{ll}
%\tilde{s}^{-}_{th,j} = & s^{-}_{j} - \pi a_{j} \Big[ \frac{1}{2} \left( 1 + a_{j} \right) B_{\nu}(T_{j}) 
%                                                                   + \frac{1}{2} \left( 1 - a_{j} \right) B_{\nu}(T_{j-1}) \\ &
%                                                                   + \Big(B_{\nu}(T_{j}) - B_{\nu}(T_{j-1})\Big) \left(1 - a_{j} + \frac{a_{j}}{\ln\left(1-a_{j}\right)}\right) \Big] \ ,
%\tilde{s}^{-}_{th,j} = & s^{-}_{j} - \pi \Big[ a_{j} B_{\nu}(T_{j+1}) +  \Big(B_{\nu}(T_{j}) - B_{\nu}(T_{j-1})\Big) \left(1 - a_{j}\right) \\ &
%                                                                - a_{j} \frac{B_{\nu}(T_{j}) - B_{\nu}(T_{j-1}}{\ln\left(1-a_{j}\right)} \Big]
\tilde{s}^{-}_{th,j} =  s^{-}_{j} - \pi \Big[ a_{j} B_{\nu}(T_{j+1}) +  \Big(B_{\nu}(T_{j}) - B_{\nu}(T_{j-1})\Big) \left(1 - a_{j}\right) 
                                                                + a_{j} \frac{B_{\nu}(T_{j}) - B_{\nu}(T_{j-1})}{\ln\left(1-a_{j}\right)} \Big] \ ,
%\end{array}
\label{eqn:source_scaled_thdn}
\end{equation}
\normalsize
where $\tilde{s}^{+}_{th,j}$ and $\tilde{s}^{-}_{th,j}$ are the adjusted layer thermal source 
terms.  { These adjustments derive from \citep{cloughetal1992} by replacing their transmissivity 
($T$ in their notation) with $1-a_{\nu}$ (i.e., one minus the absorptivity), and by representing the 
layer optical depth in \citep{cloughetal1992} as the log of the layer transmissivity 
[or $-\ln\left(1-a_{j}\right)$], following the definition of optical depth.} Our expressions give 
Planck-like contributions that go to $a_{j} \pi \left( B_{\nu}(T_j) + B_{\nu}(T_{j-1}) \right)/2$ for 
small $a_{j}$.  And, in the opposite extreme, where $a_{j} \rightarrow 1$, the Planck-like 
contributions ensure that the difference between the upwelling and downwelling layer source 
terms (which gives the net thermal flux in the optically thick limit) maintain information 
about the temperature gradient and, thus, behaves like radiation diffusion, with  
$\left(B_{\nu}(T_{j}) - B_{\nu}(T_{j-1})\right)/\ln\left(1-a_{j}\right)$ $\sim$ $\Delta B_{\nu}/\ln\left(t_{j}\right)$ 
$\sim$ $\Delta B_{\nu}/\Delta \tau_{\nu}$.]  

%Recall from Equation~\ref{eqn:example} that we are adapting the radiative flux profiles to changes 
%in the atmospheric and surface state using a linearized approach.  The strong temperature 
%dependence (through the Planck function) of the layer thermal source terms introduces a much more 
%non-linear response than dependences on other state properties (e.g., through temperature-dependent 
%line profiles or gas mixing ratios).  The adjusted thermal source terms in 
%Equations~\ref{eqn:source_scaled_thup} and \ref{eqn:source_scaled_thdn} serve to remove as much 
%of the Planck-derived non-linearity in the layer sources as possible, thereby substantially increasing the 
%linearity of the response of radiances to temperatures.

When considering solar sources, it is convenient to scale the layer source terms by the downwelling 
direct solar flux at the top of the atmosphere, as this flux is the source of the diffuse flux generated by 
all deeper layers.  Additionally, this removes the { wavelength-dependent} structure { due to the 
solar source} from the layer source terms, making them smoother functions of wavelength.  The 
adjusted solar layer source terms are dimensionless, and are given by
\begin{equation}
\tilde{s}^{+/-}_{\odot,j} = \frac{s^{+/-}_{j}}{F^{-,di}_{0}} \ .
\label{eqn:source_scaled_sol}
\end{equation}
Additionally, we define a direct transmissivity for the solar beam, given by
\begin{equation}
t^{di}_{j} = \frac{F^{-,di}_{j}}{F^{-,di}_{0}} \ .
\label{eqn:direct_tj}
\end{equation}

%%%
\subsection{Derivatives of Layer Radiative Properties}
\label{sec:propderiv}
%%%

Using the full-physics model, we can determine how the layer radiative properties change 
given a change in the atmospheric and surface state.  The combination of the layer radiative 
properties for the original atmospheric state and their derivatives then allow us 
to linearly adapt the radiative flux profiles to a change in the state by using the flux adding 
technique.

Radiances are generally known to be a strongly non-linear function of the atmospheric and surface 
state and, thus, it may seem counter-intuitive to describe their response using linear theory.  However, 
several aspects of our approach help to limit non-linearity.  First, we resolve and correct the 
solar and thermal fluxes on a relatively high resolution spectral grid (i.e., 1--10~cm$^{-1}$).  
Second, and as outlined above, we have introduced analytic corrections to remove the strong 
temperature sensitivity of layer thermal source terms through their dependence on the Planck 
function, which also serves to improve linearity.  Finally, we limit the application of our (linear) 
Jacobians to small changes in the atmospheric and surface state.

{ To calculate the Jacobians for the layer radiative properties, our full-physics model currently 
approximates the derivatives as finite differences.  We recognize that these Jacobians could 
be derived analytically, and that this approach would improve their accuracy and speed 
\citep{spurr&christi2007}. Finite differences were adopted here to facilitate the use of the 
widely-used {\tt DISORT} algorithm \citep{stamnesetal1988} to solve the radiative transfer equation 
and expedite the development and testing of the novel features of the LiFE concept. A 
full-physics forward model that computes analytic Jacobians is being developed in parallel.}

{ To define the finite difference Jacobians for the layer radiative properties, the full-physics 
model is used to derive frequency-dependent radiances, fluxes, and associated radiative 
properties for a background atmospheric and surface state.  The layer radiative properties 
are then re-derived for an atmospheric and surface state with perturbations to any combination 
of atmospheric temperatures and optical properties (layer optical depths, single scattering 
albedos, scattering phase functions), and/or surface temperatures and albedos.}  If the 
difference between the perturbed layer transmissivity and the original transmissivity is found 
to be $\Delta t_{j}$, we then have
\begin{equation}
\frac{\partial t_{j}}{\partial \qvec{x}_{m}} \approx \frac{\Delta t_{j}}{\Delta \qvec{x}_{m,j}} \ ,
\label{eqn:partial_defn}
\end{equation}
where $\Delta \qvec{x}_{m,j}$ is the perturbed temperature or optical property, as well as 
similar relations for the layer reflectivities and source terms.  Note that each 
Jacobian has dimensions equal to the number of spectral grid points  by the number of 
atmospheric layers.  { Jacobians with respect to physical atmospheric quantities like gas or 
condensate mixing ratios are determined via a simple application of the chain rule using the 
Jacobians that were derived for all dependent optical properties.}

%Such partial 
%derivatives can be computed for a collection of dynamic elements in the atmospheric 
%state.  These form a Jacobian for each layer radiative property with respect to each dynamic element of 
%the state vector,

%To determine the Jacobians for the layer radiative properties, we perturb any variable component 
%of the atmospheric and surface state vector, $\qvec{x}_{m}$, by a small amount for each atmospheric 
%layer, $\Delta \qvec{x}_{m,j}$, and then use the full-physics model to determine the layer radiative 
%properties for the perturbed state.  If the difference between 

For example, the dynamic elements of the atmospheric state vector may be the 
temperature profile and a mixing ratio profile for a condensible species.  { A temperature 
change will affect the source function and the optical depth and single scattering albedo profiles 
(through temperature-dependent changes in the gas absorption and/or aerosol optical 
cross-sections). A change in gas mixing ratio will affect all of the layer optical properties.
To compute the partial derivatives of the layer radiative properties with respect to these elements, 
we would first perturb the temperature by a small amount (e.g., a few percent) in each atmospheric 
layer.  We then run a full-physics calculation to find the new transmissivities, reflectivities, and 
source terms, and use these to compute the temperature Jacobian.  The temperature profile is then 
reset to its original value, and the optical property profiles would each be perturbed.  Re-running the 
full physics calculation for the perturbed optical properties then gives us the partial derivatives of 
the layer radiative properties with respect to the condensible species' mixing ratio (again, via an 
application of the chain rule), thus forming a mixing ratio Jacobian.  By computing Jacobians 
with respect to optical properties and applying the chain rule to translate these into mixing ratio 
Jacobians we avoid needing to re-run the full-physics model for every variable gas or aerosol 
element.}

For a new atmospheric state where a collection of state vector elements have changed 
from the original state by $\delta \qvec{x}_{1}$, $\delta \qvec{x}_{2}$, ...
$\delta \qvec{x}_{M}$, the change in the layer radiative properties for the new atmospheric 
state are determined using the associated Jacobians, and are given by
\begin{equation}
\begin{array}{c}
\delta t_{j} = \frac{\partial t_{j}}{\partial \qvec{x}_{1}}\delta \qvec{x}_{1,j} + 
    \frac{\partial t_{j}}{\partial \qvec{x}_{2}}\delta \qvec{x}_{2,j} + \dots
    \frac{\partial t_{j}}{\partial \qvec{x}_{M}}\delta \qvec{x}_{M,j} \\
\delta t^{di}_{j} = \frac{\partial t^{di}_{j}}{\partial \qvec{x}_{1}}\delta \qvec{x}_{1,j} + 
    \frac{\partial t^{di}_{j}}{\partial \qvec{x}_{2}}\delta \qvec{x}_{2,j} + \dots
    \frac{\partial t^{di}_{j}}{\partial \qvec{x}_{M}}\delta \qvec{x}_{M,j} \\
\delta r_{j} = \frac{\partial r_{j}}{\partial \qvec{x}_{1}}\delta \qvec{x}_{1,j} + 
    \frac{\partial r_{j}}{\partial \qvec{x}_{2}}\delta \qvec{x}_{2,j} + \dots
    \frac{\partial r_{j}}{\partial \qvec{x}_{M}}\delta \qvec{x}_{M,j} \\
\delta \tilde{s}^{+}_{j} = \frac{\partial s^{+}_{j}}{\partial \qvec{x}_{1}}\delta \qvec{x}_{1,j} + 
    \frac{\partial s^{+}_{j}}{\partial \qvec{x}_{2}}\delta \qvec{x}_{2,j} + \dots
    \frac{\partial s^{+}_{j}}{\partial \qvec{x}_{M}}\delta \qvec{x}_{M,j} \\
\delta \tilde{s}^{-}_{j} = \frac{\partial s^{-}_{j}}{\partial \qvec{x}_{1}}\delta \qvec{x}_{1,j} + 
    \frac{\partial s^{-}_{j}}{\partial \qvec{x}_{2}}\delta \qvec{x}_{2,j} + \dots
    \frac{\partial s^{-}_{j}}{\partial \qvec{x}_{M}}\delta \qvec{x}_{M,j} \\
\end{array}
\label{eqn:jacob_update}
\end{equation}
Note that the Jacobians are computed for the adjusted layer source terms, 
$\tilde{s}^{+/-}$, and any change in the un-adjusted layer source terms is 
determined using Equations~\ref{eqn:source_scaled_thup}, \ref{eqn:source_scaled_thdn}, 
and \ref{eqn:source_scaled_sol}.

%%%
\subsection{Summary of the LiFE Approach}
%%%

To summarize the LiFE approach, we begin with an initial atmospheric state, $\qvec{x}$, and:
\begin{enumerate}

\item use a full-physics model to determine frequency-dependent profiles of 
      solar downwelling direct flux and upwelling and downwelling diffuse flux;

\item use a full-physics model to determine frequency-dependent profiles of 
      upwelling and downwelling diffuse thermal flux;
      
\item use a full-physics model to determine the frequency-dependent layer 
      diffuse flux transmissivity and reflectivity (Equations~\ref{eqn:defn_tj} 
      and \ref{eqn:defn_rj});
      
\item use the solar downwelling direct flux profile to determine the 
      frequency-dependent direct flux transmissivity 
      (Equation~\ref{eqn:direct_tj});
      
\item use the diffuse flux profiles and the layer transmissivities and 
      reflectivities to determine the frequency-dependent layer source
      terms (Equations~\ref{eqn:layer_sup} and \ref{eqn:layer_sdn}), and the 
      adjusted layer source terms (Equations~\ref{eqn:source_scaled_thup}, 
      \ref{eqn:source_scaled_thdn}, and \ref{eqn:source_scaled_sol}).

\end{enumerate}

Then, for all dynamic elements of the state vector, we: 

\begin{enumerate}

\item perturb the dynamic element in each atmospheric layer by a small 
      amount, $\Delta \qvec{x}_{m,j}$;
      
\item use the full-physics model to repeat the steps above, determining 
      $\Delta t_{j}$, $\Delta r_{j}$, and $\Delta t^{di}_{j}$, as well as 
      $\Delta \tilde{s}^{+/-}_{j}$ for solar and thermal sources;
      
\item use the changes in the layer radiative properties to construct a 
      Jacobian for each property for the dynamic element of the state vector, 
      which is a matrix of partial derivatives of dimensions equal to the 
      number of spectral bins by the number of atmospheric layers 
      (e.g., Equation~\ref{eqn:partial_defn}).

\end{enumerate}

Finally, once the atmospheric state has evolved to a new state, 
$\qvec{x} + \delta\qvec{x}$, the radiative flux profiles can be updated by:

\begin{enumerate}

\item using the Jacobians to compute the layer radiative properties for the 
      new atmospheric state (expressions in Equation~\ref{eqn:jacob_update}, and 
      using Equations~\ref{eqn:source_scaled_thup}, \ref{eqn:source_scaled_thdn}, 
      and \ref{eqn:source_scaled_sol} to solve for the un-adjusted source terms);
      
\item using the updated direct flux transmissivity and Equation~\ref{eqn:direct_tj} 
      to solve for the new frequency-dependent direct solar flux profile
      
\item using the updated layer radiative properties and the flux adding scheme 
      (outlined in Section~\ref{sec:flxadding}) to compute the new upwelling 
      and downwelling, frequency-dependent diffuse solar and thermal flux profiles.

\end{enumerate}

%%%
\section{LiFE Validations}
\label{sec:validation}
%%%

\begin{figure}
  \centering
  \includegraphics[scale=0.8]{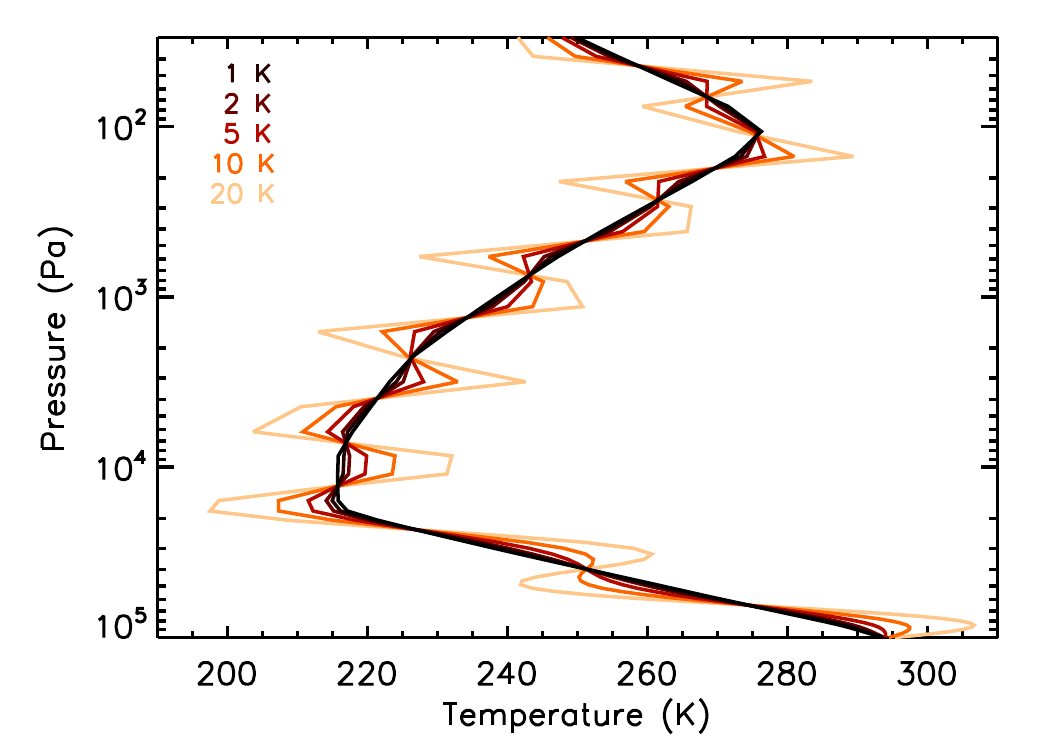}
  
  \caption{{ Standard (black) and perturbed (red) temperature profiles used in our validation 
          experiment.  Temperature perturbations vary sinusoidally in altitude with an amplitude indicated by 
          shade.}}
  \label{fig:demo_Tp}
\end{figure}

\begin{figure}
  \centering
  \includegraphics[scale=0.8]{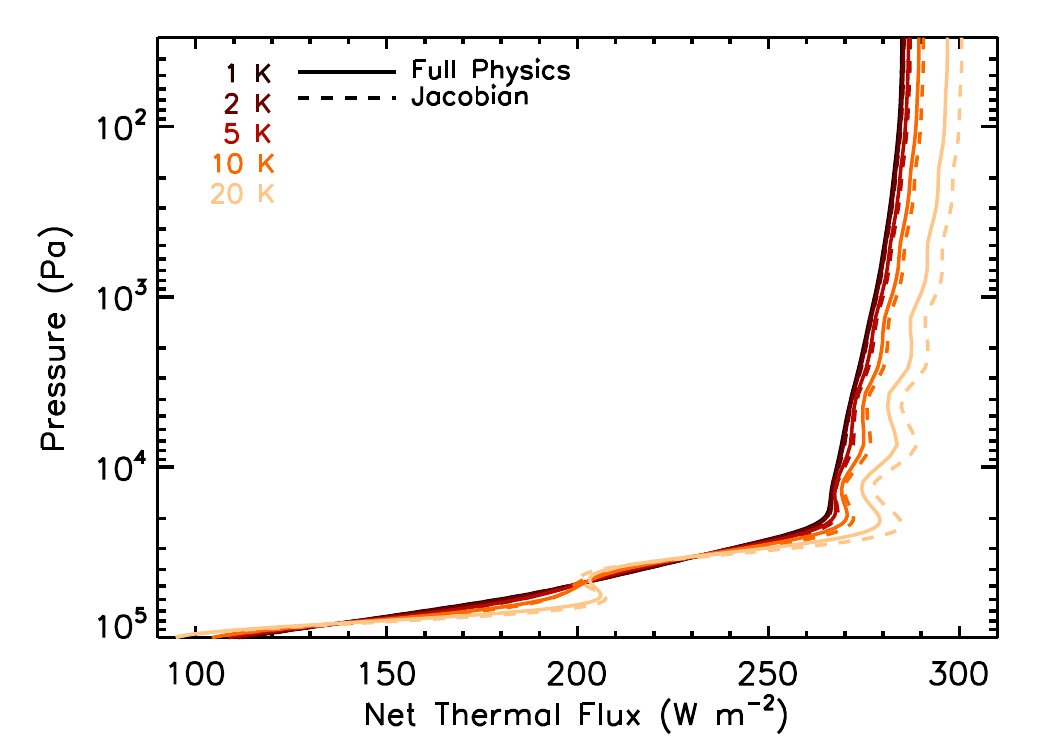}
  
  \caption{{ Net thermal flux profiles computed for perturbed temperature profiles (Figure~\ref{fig:demo_Tp}) 
         using a full-physics radiative transfer model (solid) and our Jacobian-based method (dashed).}
  \label{fig:demo_fluxes}}
\end{figure}

\begin{figure}
  \centering
  \begin{tabular}{cc}
    \includegraphics[scale=0.38]{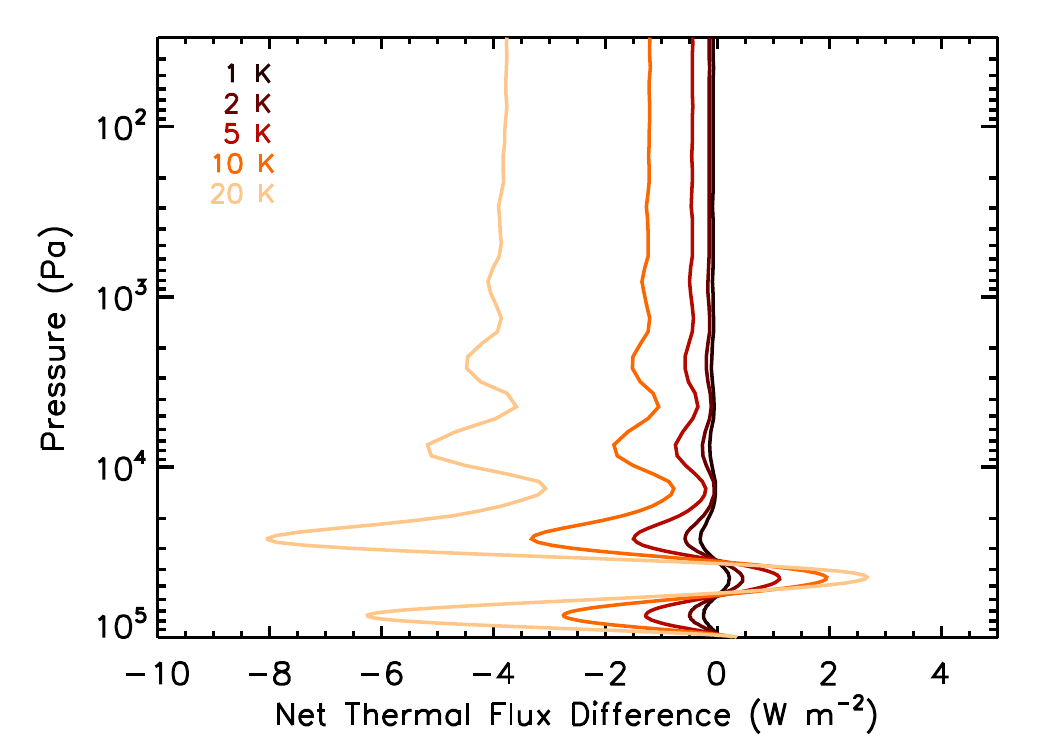} &
    \includegraphics[scale=0.38]{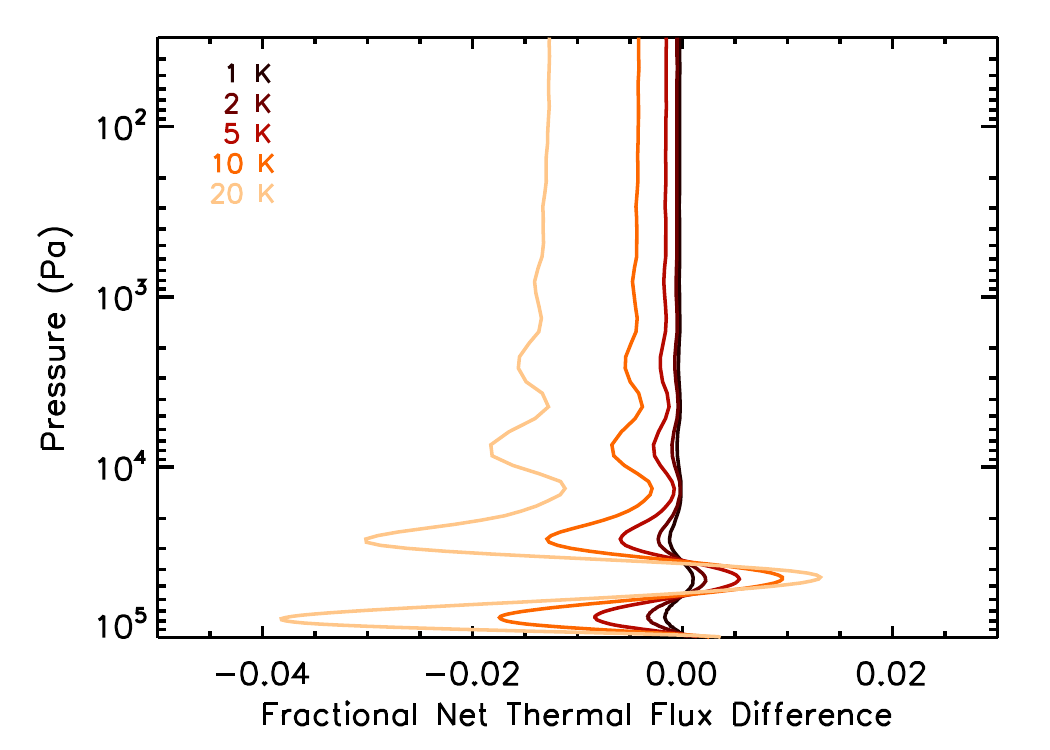} \\
  \end{tabular}
  
  \caption{{ Absolute (left) and fractional (right) differences in net thermal flux profiles computed for 
         perturbed temperature profiles (Figure~\ref{fig:demo_Tp}) when comparing our full-physics 
         radiative transfer model and our Jacobian-based method.}}
  \label{fig:demo_diffs}
\end{figure}

{
To explore the accuracy of the Jacobian-based approach of LiFE, we perform an experiment 
where we begin with a standard atmospheric state, compute solar and thermal Jacobians 
for temperature, and then compare the LiFE-derived radiative flux profiles to those from a 
full-physics tool for a number of perturbed atmospheric states.  The full-physics tool used in these 
validations is the Spectral Mapping Atmospheric Radiative Transfer ({\tt SMART}) model 
\citep[developed by D. Crisp, see][]{meadows&crisp1996}, which is a one-dimensional, multiple 
scattering, line-by-line (LBL) radiative transfer model.  The {\tt SMART} model uses a 
well-documented and stable discrete ordinate algorithm \citep[{\tt DISORT},][]{stamnesetal1988} to 
solve the radiative transfer equation.  Importantly, the {\tt SMART} model is a well-validated tool 
\citep{crisp97,savijarvietal2005,halthoreetal2005,robinsonetal2011}, and can be used to generate 
radiance spectra and flux profiles in vertically inhomogeneous, non-isothermal, plane-parallel 
scattering, absorbing, and emitting planetary atmospheres.

The standard atmosphere we adopt for our validations is a commonly-used atmospheric model 
for Earth \citep{mcclatcheyetal1972}.  We include absorption opacity due to water vapor, carbon 
dioxide, and ozone.  Perturbed temperature profiles are generated by adding a sinusoidally-varying 
component to the standard profile whose periodicity (in altitude) is 7.5~km and whose amplitude 
is either 1, 2, 5, 10, or 20~K.  The standard and perturbed temperature profiles are shown in 
Figure~\ref{fig:demo_Tp}.

Figure~\ref{fig:demo_fluxes} shows the net thermal flux profiles computed for our perturbed 
temperature profiles using both our full-physics model as well as our Jacobian approach.  
Solar fluxes are largely insensitive to temperature, so are not shown here.  For small temperature 
perturbations (i.e., less than or equal to 5~K), the flux profiles computed via our two different 
approaches are largely indistinguishable.  Figure~\ref{fig:demo_diffs} shows the absolute and 
relative differences in the net thermal flux profiles when comparing the full-physics or 
Jacobian-based approaches.  Even for 20~K temperature perturbations, 
the Jacobian-based approach reproduces the full-physics profiles to within 4\%.  Thus, by removing 
the Planck-like contribution to our layer thermal source terms, our Jacobian-based approach remains 
accurate despite large temperature perturbations.
}

%%%
\section{Example Applications of the LiFE Approach}
\label{sec:examples}
%%%

In this section we provide three demonstrations, each with increasing complexity, and each touching 
on a different planet in the Solar System.  First, we compute the wavelength-dependent layer radiative 
properties and their temperature Jacobians for a standard Martian atmosphere.  This provides insight 
into how the layer radiative properties depend on gas and aerosol opacity sources, and how these 
properties respond to changes in atmospheric temperature.  The second demonstration uses the 
LiFE approach to time-step an Earth-like atmosphere to radiative equilibrium, thus providing an example 
of how the approach can be used to determine atmospheric thermal structure.  The final demonstration 
uses the LiFE approach and a model of atmospheric convection to time-step a standard Venus 
atmosphere to radiative-convective equilibrium.
  
In all cases below, we adopt the {\tt SMART} model (described above) as our full-physics radiative 
transfer model.  All instances of this tool are run at a sufficiently small wavenumber increments (typically less 
than 0.01~cm$^{-1}$) to resolve 
all relevant spectral lines between 50 to 10$^{5}$~cm$^{-1}$ (i.e., 0.1 to 200~$\mu$m), and, to limit runtime, 
the resulting layer absorptivities, transmissivities, and source terms are averaged over 5~cm$^{-1}$ intervals 
in the thermal infrared and 10~cm$^{-1}$ intervals in the visible.  Depending on the complexity of the model 
atmosphere and the desired number of Jacobians, our full-physics model requires hours to tens of hours to 
generate the requisite outputs, while subsequently using the LiFE approach to adapt the radiative fluxes 
towards equilibrium only takes seconds.

%%%
\subsection{Mars: Layer Radiative Properties and Temperature Jacobians}
\label{sec:mars}
%%%

We use a standard collection of planetary-average atmospheric properties for Mars to compute 
the layer radiative properties and their Jacobians.  We adopt a temperature profile from 
\citep{conrathetal1973} and a dust optical depth profile from \citep{conrath1975}.
%Figure~\ref{fig:mars_temp} shows our assumed temperature profiles 
%\citep[from][]{conrathetal1973}, Figure~\ref{fig:mars_tau} shows 
%the dust optical depth profile at 0.6~$\mu$m \citep[based on][]{conrath1975}, and our assumed  
%dust single-scattering albedo is shown in Figure~\ref{fig:mars_ssalb}.  
For simplicity, we assume that the only radiatively active gas is CO$_{2}$, which is taken to 
have a mass mixing ratio of 0.95.  Using the {\tt SMART} model, we generate the layer radiative 
properties, separating solar and thermal sources, and we also numerically evaluated temperature 
Jacobians for these properties, assuming a 1\% change in temperature at each atmospheric level.

\begin{figure}
  \centering
  \includegraphics[scale=0.8]{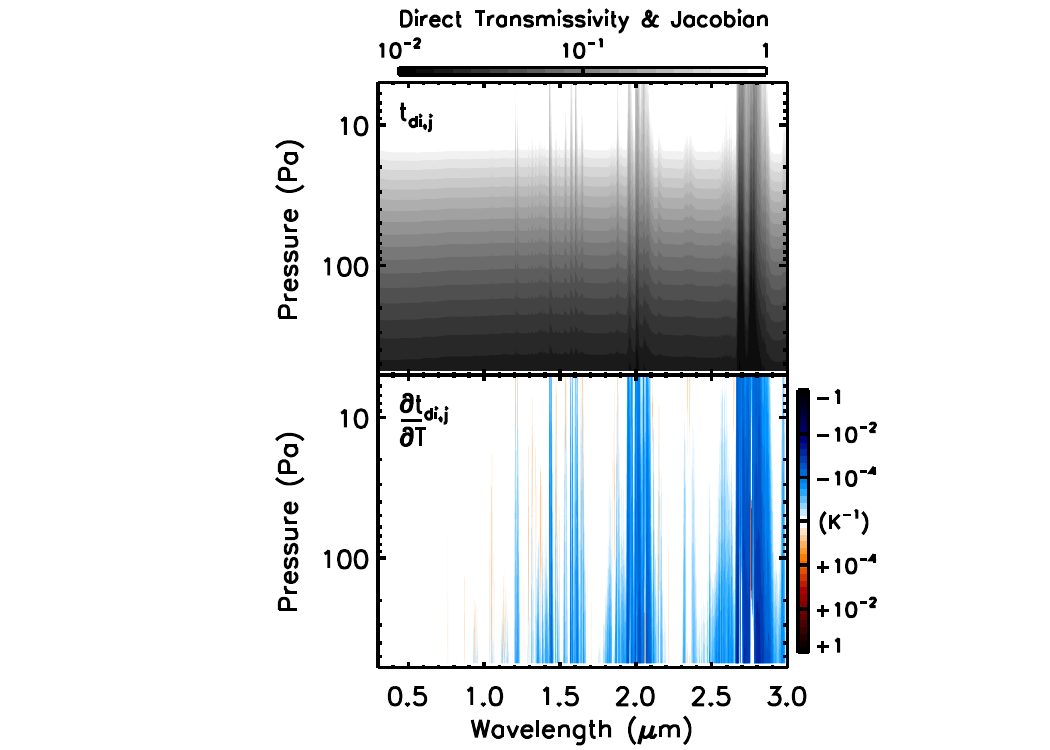}
  
  \caption{Wavelength-dependent transmissivity for the direct solar beam in
	the Martian atmosphere (top) and its temperature Jacobian (bottom).  For the transmissivity, 
	darker shades indicate low transmission.  For the Jacobians, red shades indicate increasing 
	transmissivity with increasing temperature, and blue shades indicate decreasing transmissivity 
	for increasing temperature.}
  \label{fig:mars_sol_dir}
\end{figure}

Figure~\ref{fig:mars_sol_dir} shows the transmissivity for the direct solar flux and the 
associated temperature Jacobian as shaded contours in wavelength and atmospheric 
pressure.  The transmissivity is dominated by dust opacity at most wavelengths, but 
CO$_{2}$ absorption bands can clearly be distinguished.  The Jacobians indicate that the 
transmissivities are only weakly sensitive to temperature at these wavelengths.  An increase 
in temperature leads to weaker absorption at the centers of absorption bands, and stronger 
absorption in the wings.  This behavior is due to the temperature dependence in the individual 
line half-widths (which are predominately Doppler broadened here), and the Boltzmann-like 
temperature dependence in the line strengths, both of which lead to decreased opacity 
(increased transmissivity) near band centers for increased temperature.  Note that absorption in 
the centers of some CO$_{2}$ bands are so strong in the deepest portions of the atmosphere 
that their transmissivity is zero for all temperatures used in these models, and so the 
Jacobians show no temperature sensitivity here.

\begin{figure}
  \centering
  \includegraphics[scale=0.8]{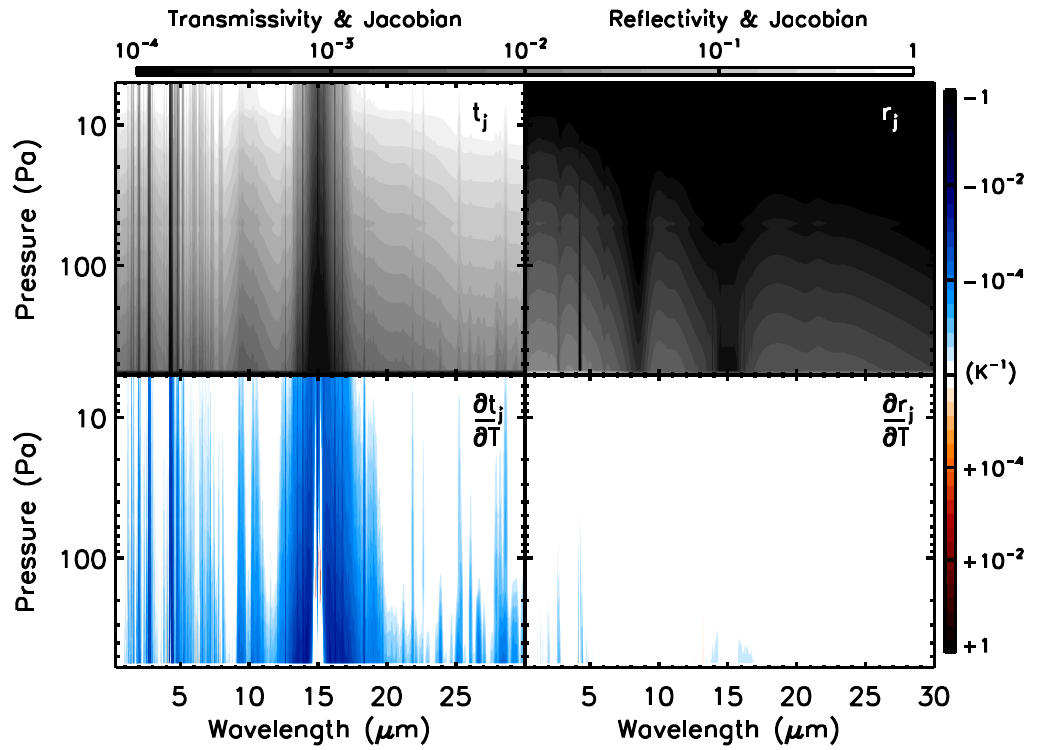}
  
  \caption{Wavelength-dependent layer diffuse flux transmissivity (top-left) and reflectivity (top-right) in 
  	the Martian atmosphere and their temperature Jacobians (bottom).  For the transmissivity and its 
	Jacobian, shading is the same as in Figure~\ref{fig:mars_sol_dir}.  For the reflectivity, lighter shades 
	indicate higher layer reflectance, and, for the Jacobians, red shades indicate increasing reflectivity with 
	increasing temperature, while blue shades indicate decreasing reflectivity for increasing temperature.}
  \label{fig:mars_transrefl}
\end{figure}

Layer diffuse flux transmissivity and reflectivity, as well as their associated temperature 
Jacobians, are shown in Figure~\ref{fig:mars_transrefl}.  As was the case for the 
direct solar flux, the transmissivity is dominated by dust opacity except in the CO$_{2}$ 
absorption bands, with the 4.3~$\mu$m and 15~$\mu$m bands being especially absorptive.  Layer 
reflectivity is generally small, and drops to zero at the centers of strong CO$_{2}$ features.  
Like the direct solar terms, the transmissivity Jacobians show that increased temperatures cause 
decreased absorption near band centers, and increased absorption in band wings.  Due to the 
increased absorption in band wings at higher temperatures, the reflectivity decreases at these 
wavelengths as temperature increases.

\begin{figure}
  \centering
  \includegraphics[scale=0.8]{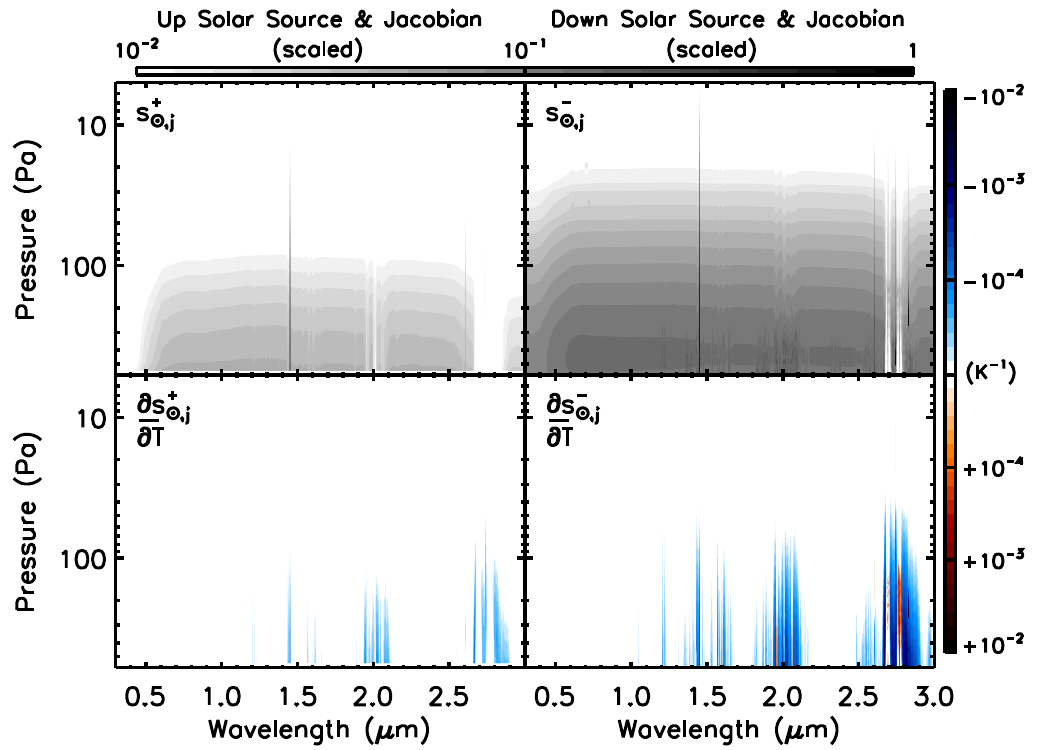}
  
  \caption{Wavelength-dependent upwelling (left) and downwelling (right) layer solar source terms (top) and 
  	their temperature derivatives (bottom) for Mars.  Both the source terms and their derivatives have been 
	scaled by (1) the top-of-atmosphere solar flux, which removes wavelength-dependent structure from the 
	solar spectrum, and (2) $d\ln p_{j}$, which removes layer path length effects and creates source terms 
	that are smooth functions of pressure.  For the source terms, darker shades indicate larger flux sources.  
	For the Jacobians, red shades indicate increasing sources with increasing temperature, and blue shades 
	indicate decreasing sources with increasing temperature.}
  \label{fig:mars_sol_source}
\end{figure}

Figure~\ref{fig:mars_sol_source} shows the layer solar source terms, which have been adjusted by 
the top-of-atmosphere solar flux to remove wavelength-dependent structure resulting from the 
solar spectrum.  As the dust particles are forward-scattering, the downwelling solar source 
terms tend to be larger than the upwelling source terms.  Wavelength-dependent structure in 
both the upwelling and downwelling source terms is due primarily to the dust optical properties, 
which have a lower single-scattering albedo below about 0.6~$\mu$m.  
Note that the source terms are smaller (or vanishingly small) within 
CO$_{2}$ absorption features, since the layers strongly absorb sunlight at these wavelengths, 
rather then scattering it into the diffuse radiation field.  Finally, the temperature Jacobians 
show a decrease in the source terms for an increase in temperature in the wings of CO$_{2}$ 
features, and an increase in the cores.  This is due to the decrease in transmissivity 
in band wings at higher temperatures, and the increase in transmissivity in the band cores.

\begin{figure}
  \centering
  \includegraphics[scale=0.8]{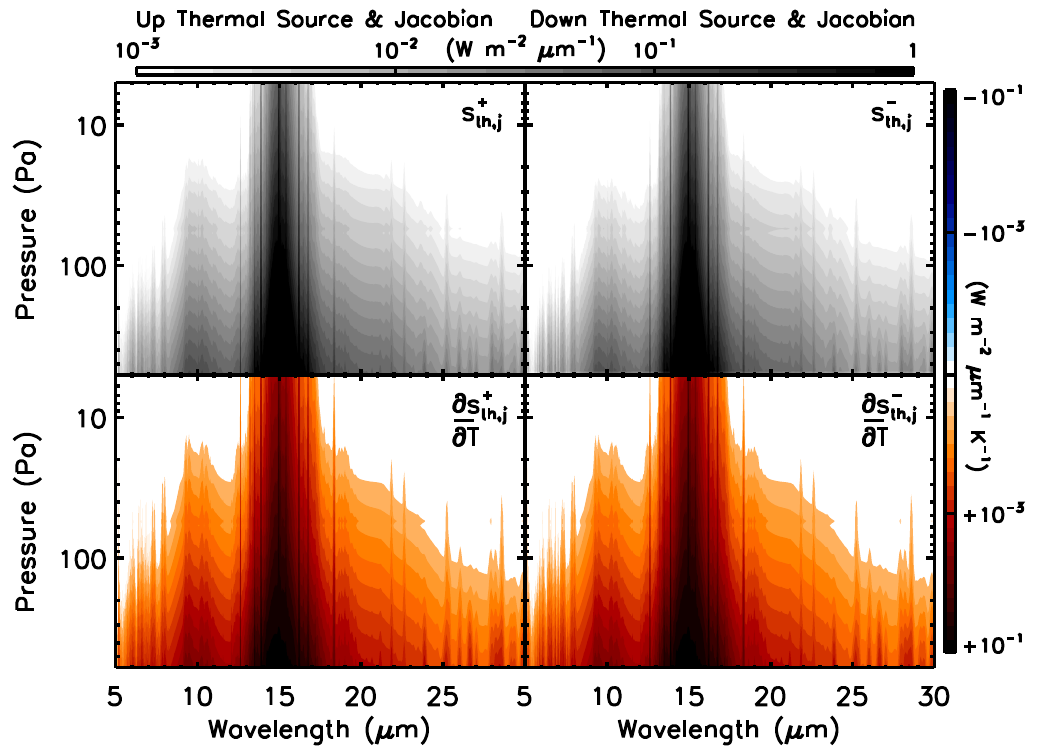}
  
  \caption{Wavelength-dependent upwelling (left) and downwelling (right) layer thermal source terms (top) and 
  	their temperature derivatives (bottom) for Mars.  For the source terms, darker shades indicate larger flux 
	sources.  For the Jacobians, red shades indicate increasing sources with increasing temperature, and blue 
	shades indicate decreasing sources with increasing temperature.}
  \label{fig:mars_therm_source}
\end{figure}

Layer thermal sources are shown in Figure~\ref{fig:mars_therm_source}.  The upwelling and downwelling 
terms are identical since the layers are equally efficient at generating upwelling and downwelling 
thermal flux.  The sources are strongest in the 15~$\mu$ CO$_{2}$ band, with dust providing thermal 
radiative fluxes outside of CO$_{2}$ features.  As one might expect, the temperature Jacobians (shown 
at the bottom of this figure) demonstrate that, for an increase in temperature, the layer thermal 
source terms increase at all wavelengths.

%%%
\subsection{Earth: Using LiFE to Determine Radiative Equilibrium Thermal Structure}
\label{sec:earth}
%%%

An important aspect of the LiFE approach is that it allows the radiative fluxes in a planetary 
atmosphere to be rapidly adapted to changes in the atmospheric state, while still being guided 
by the values from the full-physics model.  Thus, the approach can be used to estimate the solar 
and thermal flux variations that occur as the thermal structure evolves from an initial state toward 
thermal equilibrium.  As a demonstration, in this section we use the LiFE approach to determine 
the radiative equilibrium thermal structure of an Earth-like atmosphere.

We assume an atmosphere that is 78\% N$_{2}$ and 21\% O$_{2}$, by volume, and we include 
H$_{2}$O, CO$_{2}$, and O$_{3}$ as trace gases.  In reality, water vapor is a condensible 
species in Earth's atmosphere, whose partial pressure is tied to atmospheric temperature.  
However, for simplicity, we hold the water vapor mixing ratio profile fixed.  For this example, we 
do not include clouds, and we assume a wavelength-dependent surface albedo appropriate for 
ocean (which has a value of ~5\% at most wavelengths).

\begin{figure}
  \centering
  \includegraphics[scale=0.8]{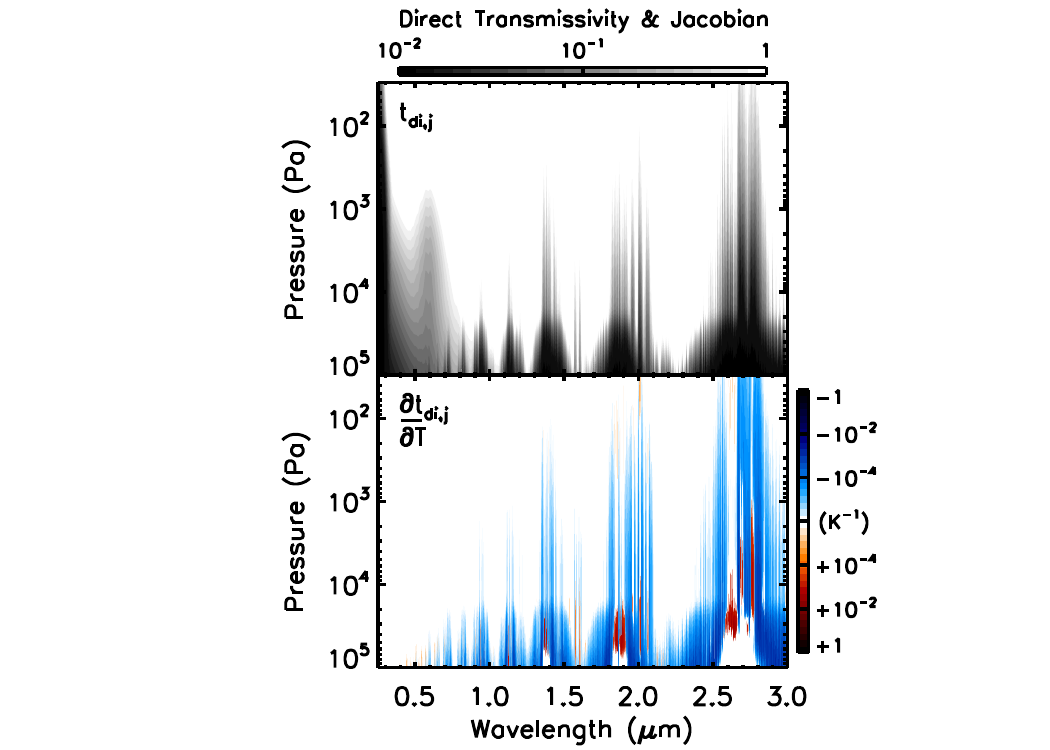}
  
  \caption{Initial wavelength-dependent transmissivity for the direct solar beam in
	Earth's atmosphere (top) and its temperature Jacobian (bottom) used in our example application 
	of the LiFE approach.  Shading is the same as Figure~\ref{fig:mars_sol_dir}.}
  \label{fig:earth_sol_dir}
\end{figure}

Direct solar transmissivity and its temperature Jacobian are shown in 
Figure~\ref{fig:earth_sol_dir}.  Strong absorption features due to water vapor can 
be seen throughout the near-infrared, contributions from CO$_2$ can be seen near 
2.2 and 2.7~$\mu$m, and opacity from ozone and Rayleigh scattering are apparent at shorter 
wavelengths.  Our extinction cross sections for ozone and Rayleigh scattering in the ultraviolet 
and visible are not temperature dependent, so these features do not appear in the temperature 
Jacobians.  The water vapor absorption bands show a similar temperature dependence to the 
absorption bands seen for Mars in the previous section.

\begin{figure}
  \centering
  \includegraphics[scale=0.8]{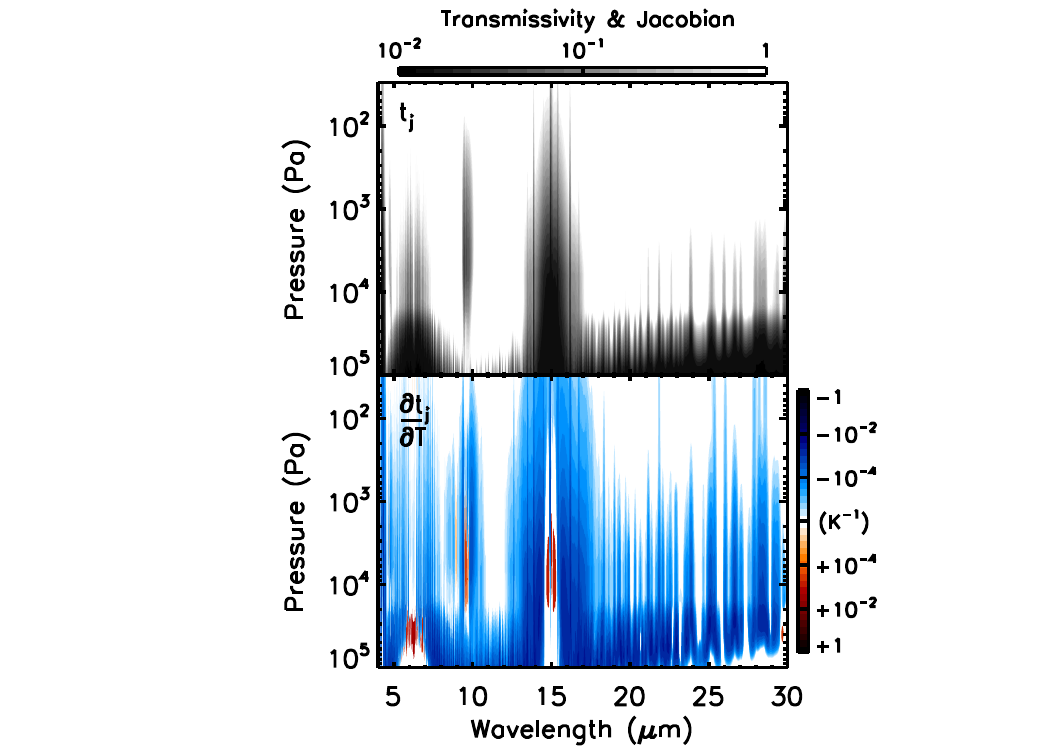}
  
  \caption{Initial wavelength-dependent layer diffuse flux transmissivity (top) in Earth's atmosphere and 
       	its temperature Jacobian (bottom) used in our example application of the LiFE approach.  Shading 
	is the same as Figure~\ref{fig:mars_transrefl}.  Layer reflectivity is not shown as reflectance is due to 
	Rayleigh scattering and, thus, is effectively zero at the wavelengths used above.}
  \label{fig:earth_transrefl}
\end{figure}

Layer transmissivity and its temperature Jacobian throughout the mid-infrared are 
shown in Figure~\ref{fig:earth_transrefl}.  Layer reflectivity is not shown as it is 
uniformly zero at these wavelengths.  Line absorption due to H$_2$O, CO$_2$, and 
O$_3$ can be clearly distinguished.  Like the Mars example, 
each band shows an increasing transmissivity with temperature at its core, and 
a decreasing transmissivity with temperature in its wings.  This effect is due, 
predominately, to the temperature dependence in line strengths.

Solar sources are shown in Figure~\ref{fig:mars_sol_source}, which are due to 
Rayleigh scattering.  Again, temperature Jacobians are not shown as the Rayleigh 
scattering extinction cross section does not depend on temperature in our model.  
Thermal sources and their temperature Jacobians are shown in 
Figure~\ref{fig:mars_therm_source}.  As would be expected, the sources show that 
thermal flux is mostly generated in water vapor, carbon dioxide, and ozone absorption 
bands, and the flux increases with temperature.

\begin{figure}
  \centering
  \includegraphics[scale=0.8]{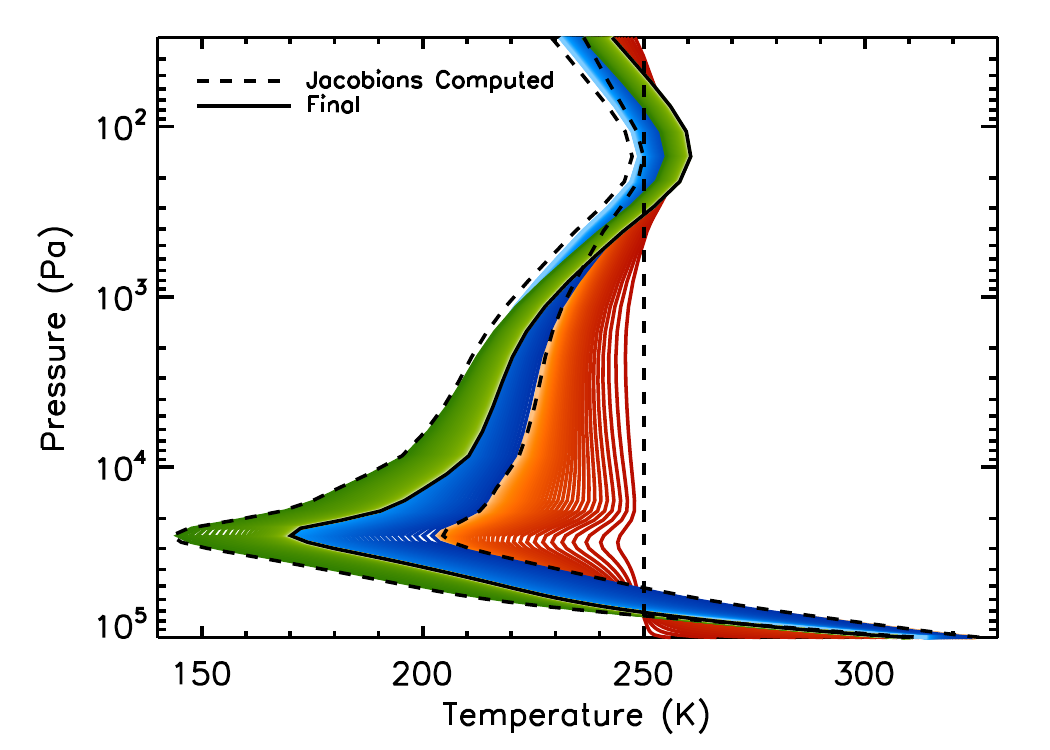}
  
  \caption{Evolution of our 1-D, cloud-free Earth atmosphere to a radiative equilibrium state, beginning 
        with an isothermal profile (dashed).  Red curves show the evolution using the first set of computed 
        layer radiative properties and their Jacobians, with lighter hues indicates later times, which ends at 
        the dashed profile.  Our full-physics model is then used to compute updated layer radiative properties 
        and Jacobians for this new atmospheric state and evolved through time, as is shown by the blue curves 
        (resulting in the temperature profile with the coldest tropopause).  This process is repeated again, 
        following the green curves with time.  The final profile is the solid line in black.}
  \label{fig:earth_tempevol}
\end{figure}

To determine the radiative equilibrium thermal structure of the atmosphere we solve the 
thermodynamic energy equation as an initial value problem.  Our calculation begins with an 
isothermal profile for which we generate the layer source terms and their temperature 
Jacobians.  We then use these to time-step the atmosphere to radiative equilibrium by 
using the radiative heating/cooling rates to update the temperature profile, and then 
using the LiFE approach to determine the radiative fluxes for the updated atmospheric 
state.  This temperature evolution of the atmosphere away from the isothermal state is 
shown by the red profiles in Figure~\ref{fig:earth_tempevol}, and the resulting 
radiative equilibrium solution is the dashed line.  Recall that this is a pure radiative equilibrium 
solution, and not a radiative-convective solution, thus resulting in a very cold tropopause.  The 
initial full-physics calculation takes about two wall clock hours on a single processor, and each 
time step using the Jacobians takes about one second.  The progression to radiative equilibrium 
from the isothermal state takes 2,000 time steps, which translates to five months of model 
time.

The solution found by application of the first set of Jacobians is not the true radiative 
equilibrium solution, since the temperature profile has evolved away from the range 
over which the Jacobians are valid.  So, we twice repeat the process outlined in the 
previous paragraph, resulting in evolution shown by the blue and green profiles in 
Figure~\ref{fig:earth_tempevol}.  Since each iteration is nearer to the true radiative 
equilibrium solution, the number of time steps required to reach equilibrium is less 
than the first iteration.  Model timescales to reach equilibrium for the second and 
thirds iterations were four months and one month, respectively.

\begin{figure}
  \centering
  \includegraphics[scale=0.8]{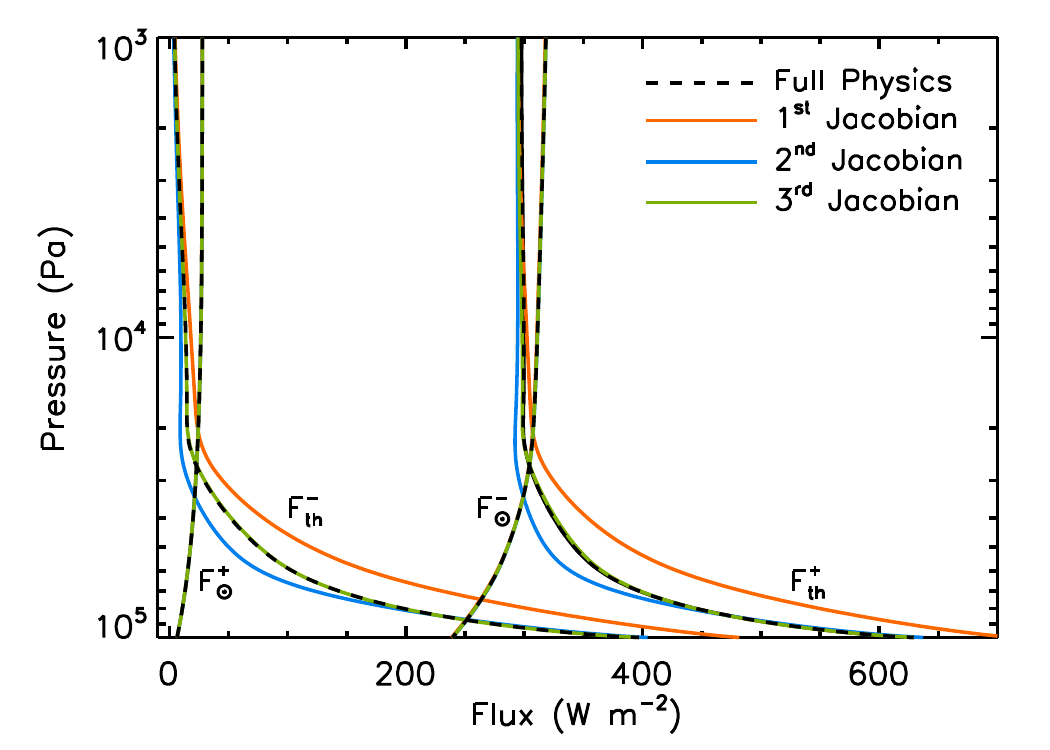}
  
  \caption{Evolution of upwelling and downwelling solar and thermal flux profiles, following 
    Figure~\ref{fig:earth_tempevol}.  Dashed lines show the radiative-equilibrium profiles, as 
    computed by the full-physics model.  Colored curves show the profiles for the different 
    radiative-equilibrium states determined by successive refinements of the layer radiative 
    properties and their Jacobians.}
  \label{fig:earth_fluxes}
\end{figure}

The accuracy of the three sets of layer radiative properties and their associated 
Jacobians in computing the radiative equilibrium solution can be seen by comparing 
the solar and thermal flux profiles to those computed by the full-physics model.  
Figure~\ref{fig:earth_fluxes} demonstrates how the different sets of radiative properties 
and Jacobians approach the true solution.  Upwelling and downwelling solar and 
thermal flux profiles are shown, with the ``true" radiative equilibrium solution (from 
the full-physics model) shown as dashed lines.  All sets of layer radiative properties 
and Jacobians do a good job of matching the solar flux profiles, which are largely 
insensitive to changes in atmospheric temperature.  However, the thermal fluxes 
from the radiative equilibrium solution determined from the first set of properties 
and Jacobians differ from the true fluxes by more than 50 W m$^{-2}$ at some 
pressures.  This difference shrinks to nearly zero through subsequent computations 
of properties and Jacobians, eventually arriving at the correct solution.

%%%
\subsection{Venus: Determining the Radiative-Convective Equilibrium State with 
                    the LiFE Approach}
\label{sec:venus}
%%%

The structure of real planetary atmospheres depends on more than just 
radiative balance---dynamical processes, such as convection, are also of 
critical importance.  To demonstrate how the LiFE approach works within a 1-D
radiative-convective scheme, we paired the {\tt SMART} model and the 
associated LiFE framework with a simple mixing-length approach to 
convection.  Here, the convective heating rate, $q_{c}$, is given by 
\begin{equation}
q_{c} = - \frac{1}{c_{p}\rho} \frac{dF_{c}}{dz} \ ,
\end{equation}
where $c_{p}$ is the atmospheric heat capacity, $\rho$ is the atmospheric 
density, and $F_{c}$ is the convective heat flux, taken as
\begin{equation}
F_{c} = - \rho c_{p} K_{H} \left( \frac{dT}{dz} + \Gamma_{ad} \right) \ ,
\end{equation}
where $K_H$ is the eddy diffusivity for heat, and $\Gamma_{ad}=g/c_p$ is 
the adiabatic lapse rate, where $g$ is the acceleration due to gravity.  The 
eddy diffusivity vanishes when the temperature profile is stable against 
convection, and is given by
\begin{equation}
K_{H} = 
  \begin{cases}
    l^{2} \left[ - \frac{g}{T} \left( \frac{dT}{dz} + \Gamma_{ad} \right) \right]^{1/2}, &  \frac{dT}{dz} > - \Gamma_{ad}  \\
    0, &  \frac{dT}{dz}  \leq - \Gamma_{ad}
  \end{cases}
\end{equation}
where $l$ is the mixing length.  We follow Blackadar~\citep{blackadar1962}, taking the 
mixing length to be
\begin{equation}
l = \frac{kz}{1 + kz/l_0} \ ,
\end{equation}
where $k$ is von K\'{a}rm\'{a}n's constant, and $l_0$ is the 
mixing length in the free atmosphere, which we take as the pressure 
scale heigh, $H=RT/g$, where $R$ is the specific gas constant.

\begin{figure}
  \centering
  \includegraphics[scale=0.8]{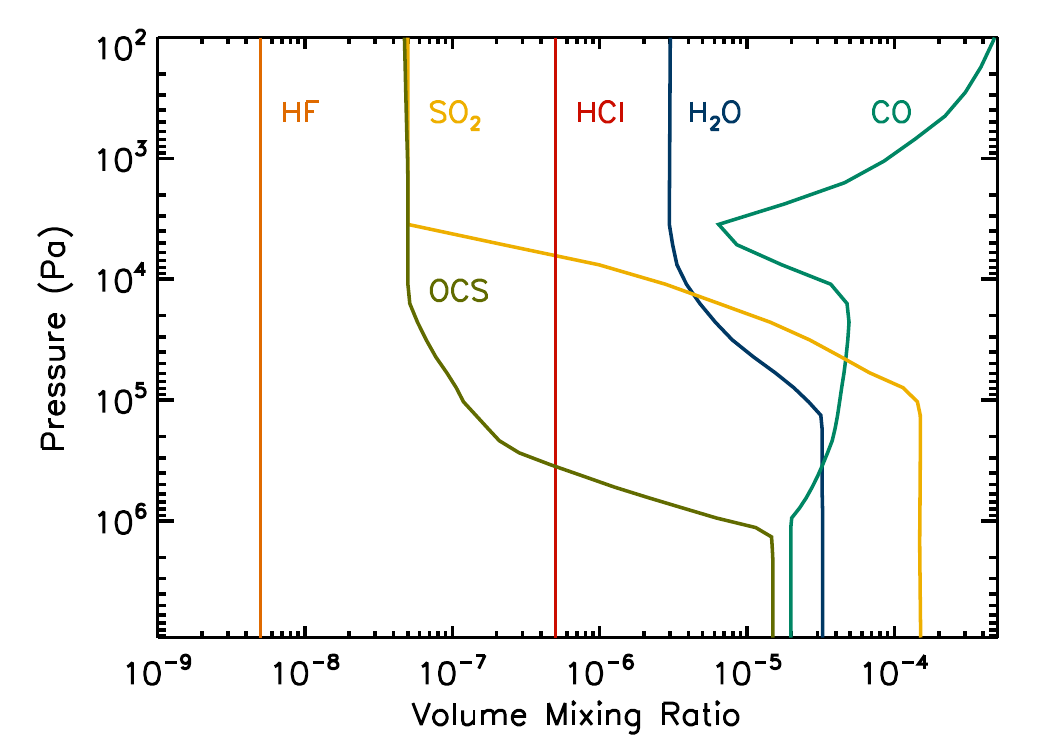}
  
  \caption{Gas mixing ratio profiles used in our Venus thermal structure calculations, from 
                Haus~et~al.~\citep{hausetal2015}.}
  \label{fig:venus_rmix}
\end{figure}

We take the atmosphere to be 96\% CO$_{2}$ and 4\% N$_{2}$, by volume, and 
we include H$_{2}$O, HDO (whose mixing ratio relative to the primary isotopologue is 
taken to be 130$\times$ larger than the telluric value), SO$_{2}$, CO, OCS, HCl, and HF 
as trace gases.  The mixing ratios for our trace gases, as well as our treatment of the 
``unknown UV absorber'', are taken from the Haus~et~al.~\citep{hausetal2015}, and gas mixing ratio 
profiles are shown in Figure~\ref{fig:venus_rmix}.  We use a standard collection of cloud 
optical properties and vertical distributions, which are taken from Crisp~\citep{crisp1986}.  Collisional 
line mixing is understood to alter CO$_{2}$ lineshapes to be sub-Lorentzian at high pressures, 
which we parameterize through a standard set of $\chi$ factors defined in 
\citep{meadows&crisp1996}.  Our treatment of collision induced absorption stems from 
a number of sources 
\citep{gruszka&borysow1997,baranovetal2004,wordsworthetal2010b,lee_yj_etal2016}.  We
use HITEMP 2010 \citep{rothmanetal2010} for our water vapor linelist, 
Huang~et~al.~\citep{huangetal2014} for our carbon dioxide linelist, and HITRAN 2012 
\citep{rothmanetal2013} for all other gases.

To decrease runtime, model time stepping began with an initial $T$-$p$ profile that had a 
surface temperature of 730 K, following a dry adiabat to 0.1 bar, and an isothermal upper 
atmosphere at 210 K.  Initial profiles further from the equilibrium solution were also found to 
approach the same solution as the case presented here, but require substantially more iterations 
of the full-physics model.  To expedite the march to equilibrium, we use a pressure-dependent 
time step, which helps account for the long thermal timescales that occur deep in the 
Venusian atmosphere.  A functional form that was found to work had $\Delta t \propto p^{1/4}$.
%\begin{equation}
%\Delta t = (4\times10^{4} \text{~s}) \left( \frac{p}{92.1 \text{~bar}} \right)^{1/4} \ .
%\end{equation}

\begin{figure}
  \centering
  \includegraphics[scale=0.8]{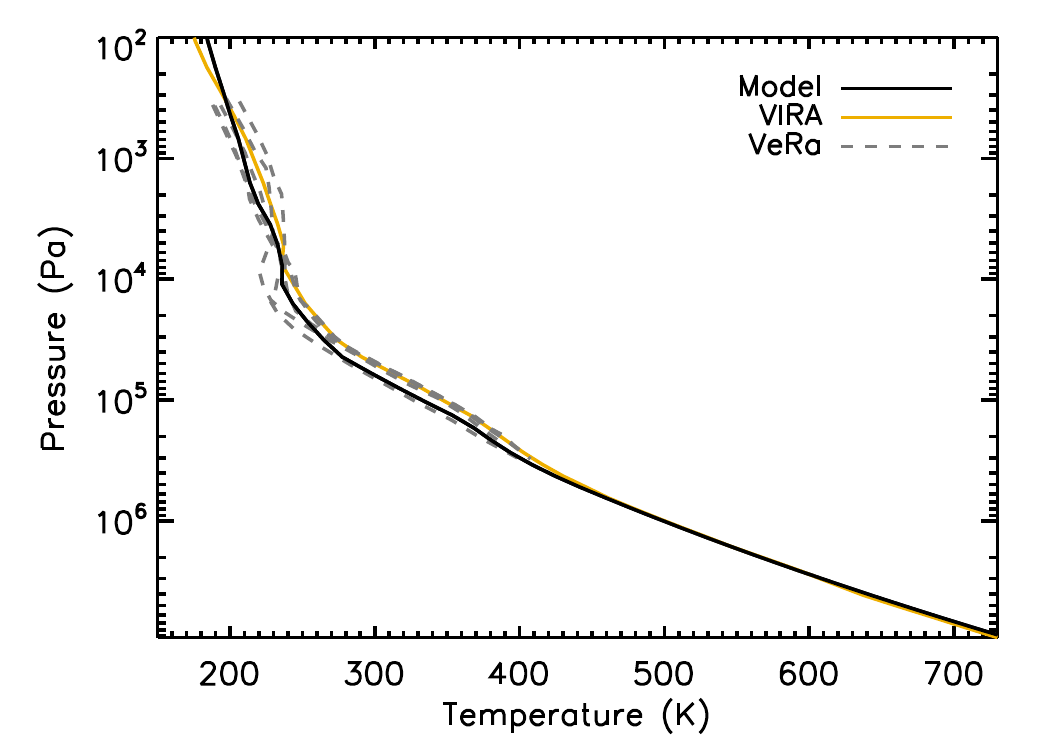}
  
  \caption{Venus radiative-convective equilibrium thermal structure from our model.  Also 
       shown is the Venus International Reference Atmosphere (VIRA) 
       \citep{moroz&zasova1997} and a number of latitude-dependent temperature profiles for 
       the upper atmosphere from the {\it Venus Express} radio science experiment (VeRa) 
       \citep{tellmannetal2009}, spanning the equatorial region to high latitudes.}
  \label{fig:venus_temp}
\end{figure}

A total of three sets of full-physics calculations were used to achieve a radiative-convective 
equilibrium state, and shortwave fluxes were computed at four Gaussian solar zenith 
angles covering the sunlit hemisphere and were combined using Gaussian quadrature 
\citep{crisp1986}.  While using multiple solar zenith angles increases runtime, this approach 
helps to ensure a better ``planetary average'' set of shortwave fluxes.  Our equilibrium 
radiative-convective profile is shown in Figure~\ref{fig:venus_temp}.  As our initial guess 
was near the final answer, the temperature evolution of the atmosphere was less dramatic 
than in the Earth case, so we only show the final $T$-$p$ profile in this figure.  Also shown 
are the Venus International Reference Atmosphere (VIRA) \citep{moroz&zasova1997} 
and measurements of the mesospheric thermal structure from {\it Venus Express} 
\citep{tellmannetal2009}.

The model reproduces the thermal structure of the Venusian atmosphere 
extremely well.  Our computed surface temperature is 736~K, which is consistent 
with the ground-truthed value of $\sim\!735$~K derived from {\it Pioneer Venus} and 
the {\it Venera} landers \citep{seiffetal1985}.  Critically, our computed mesospheric thermal 
structure is bounded by the radio occultation results from {\it Venus Express}.

\begin{figure}
  \centering
  \includegraphics[scale=0.8]{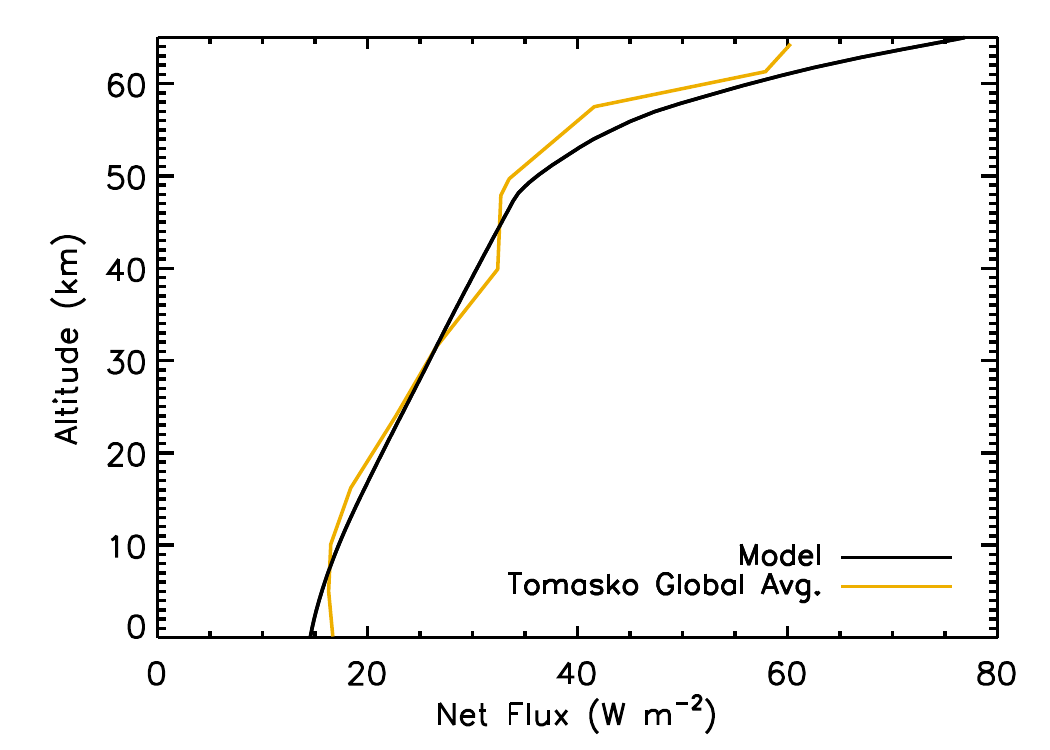}
  
  \caption{Net solar flux in the Venus atmosphere from our model (black) and from an 
                estimate of the global average from Tomasko~et~al.~\citep{tomaskoetal1980} 
                (yellow), which was based on measurements from the {\it Pioneer Venus} sounder.}
  \label{fig:venus_solflx}
\end{figure}

\begin{figure}
  \centering
  \includegraphics[scale=0.8]{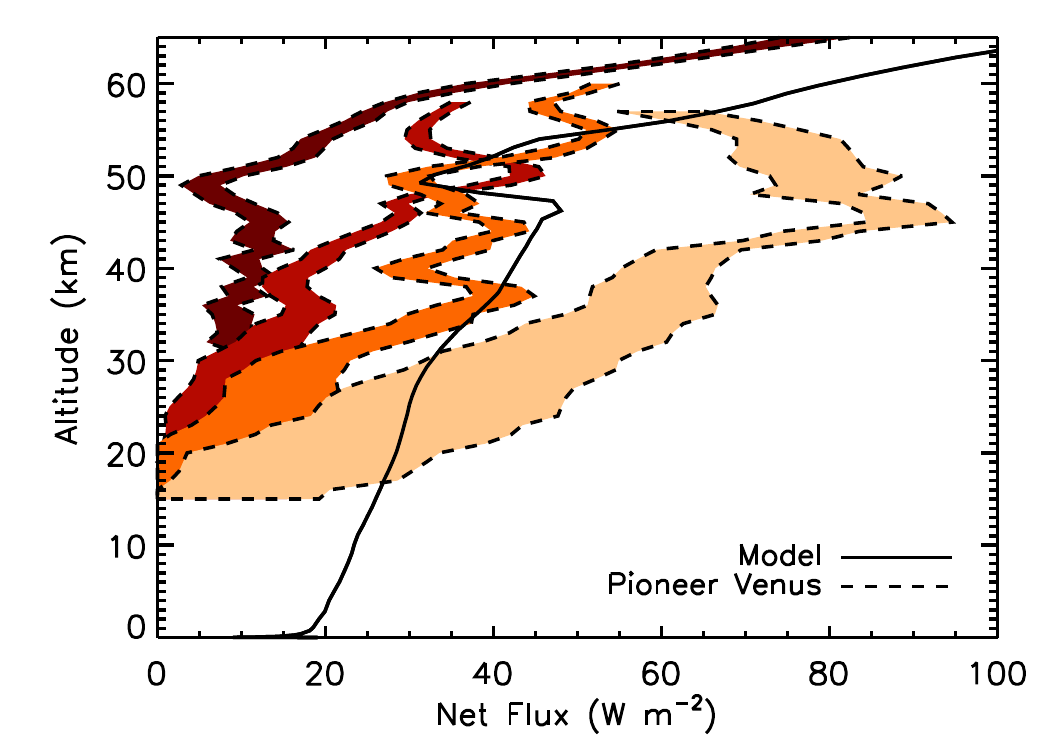}
  
  \caption{Net thermal flux in the atmosphere of Venus.  Solid line is from our model, 
       and shaded regions represent measurements and uncertainties from the {\it Pioneer 
       Venus} sounder and probes \citep[from ][]{revercombetal1985}.  From left to right 
       at 40 km, data curves are: sounder, day-side, 4$^{\circ}$ N; probe, day-side, 
       31$^{\circ}$ S; probe, night-side, 27$^{\circ}$ S; and probe, day-side, 
       60$^{\circ}$ N.}
  \label{fig:venus_irflx}
\end{figure}

Solar flux profiles from the radiative-convective equilibrium model 
are shown in Figure~\ref{fig:venus_solflx}.  Also shown is an estimate 
of the global average net flux profile, which is based on results from the 
{\it Pioneer Venus} sounder \citep{tomaskoetal1980}.  The model shortwave profiles 
are in excellent agreement with the data. Finally, the model net thermal flux profile is 
shown in Figure~\ref{fig:venus_irflx}.   Measurements and uncertainty estimates from 
the {\it Pioneer Venus} mission are also shown, and are taken from 
Revercomb~et~al.~\citep{revercombetal1985}.  In general, the model falls within the range of 
measured values, although there is an indication that the model finds a larger 
net thermal flux profile in the deepest regions of the atmosphere (i.e., below 
about 15~km).  Observations of the Venus night side 
\citep{crispetal1989,meadows&crisp1996,arneyetal2014} indicate that much of the 
variability in the {\it Pioneer Venus} net thermal fluxes may be associated with 
variations in the highly-variable thermal infrared opacity of the middle and lower clouds 
\citep{crisp&titov1997}, an effect omitted in our calculations.  The sudden decrease in net 
thermal flux near 50~km altitude is due to convection at the cloud base in our simulation.

{ 
Recently, Haus~et~al.~\citep{hausetal2017} applied a Jacobian-based technique to computing 
solar heating and thermal cooling rates in the Venusian mesosphere.  In this approach, individual 
level temperatures (or some other quantity, such as a cloud enhancement parameter) were perturbed, 
and a new net heating/cooling rate was calculated at all model layers using a full-physics model.  
Jacobians, which describe the response of a layer heating/cooling rate to a change in temperature (or 
some other parameter) at any given model level, were determined by differencing the perturbed cases 
to a baseline model.  This approach has the computational advantage of being spectrally-unresolved 
(as heating/cooling rates are integrated quantities), although a wavelength/wavenumber grid correction 
factor was required to be applied to the Jacobian-computed heating/cooling rates.  The LiFE approach, 
while spectrally-resolved, has the advantage that Jacobians are computed on {\it local} layer radiative 
properties, and perturbations to the atmosphere are then handled using radiative principles (i.e.,~a 
two-stream flux adding technique).  Temperature sensitivity is significantly reduced 
(see Section~\ref{sec:validation}) by removing a Planck-like contribution to the layer source terms 
(i.e.,~Equations~\ref{eqn:source_scaled_thup} and \ref{eqn:source_scaled_thdn}), which 
would not be the case when working with heating/cooling rate Jacobians.  Furthermore, the 
heating/cooling rate approach described by Haus~et~al. scales as the square of the number of 
atmospheric layers, whereas the LiFE approach scales linearly with the number of atmospheric 
layers (and with the number of spectral gridpoints).
}

%%%
\section{Example Model Comparison}
\label{sec:comparison}
%%%

\begin{figure}
  \centering
  \includegraphics[scale=0.8]{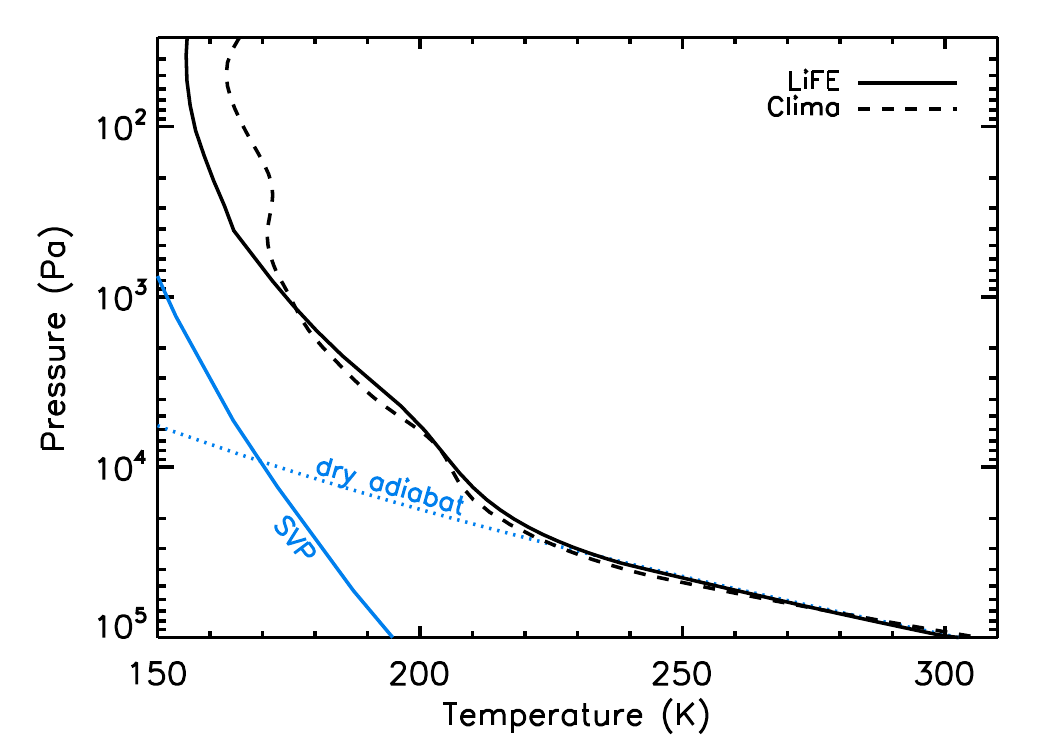}
  
  \caption{{ Comparison of the equilibrium radiative-convective thermal structures computed 
       using the {\tt Clima} model and our LiFE-based approach for a planet with a 1~bar pure 
       carbon dioxide atmosphere orbiting at 1~AU from a solar twin.  The dry adiabat and saturation 
       vapor pressure (SVP) are shown for carbon dioxide.}}
  \label{fig:climaTp}
\end{figure}

\begin{figure}
  \centering
  \includegraphics[scale=0.8]{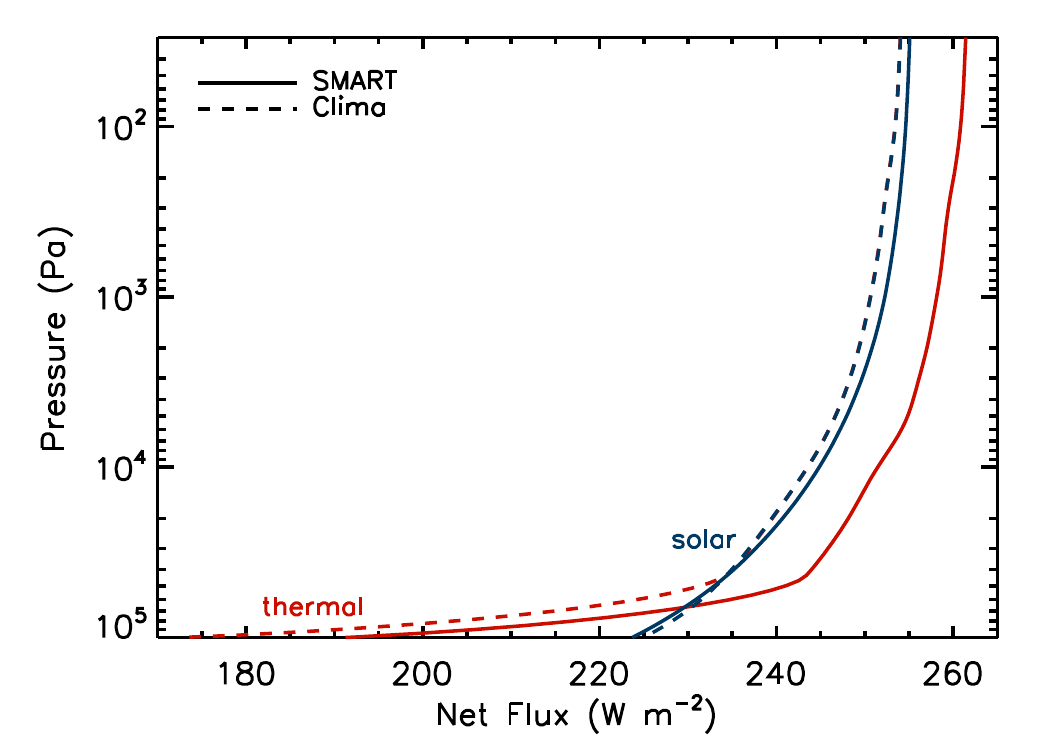}
  
  \caption{{ Net solar (blue) and thermal (red) flux profiles computed for the {\tt Clima}-derived 
       equilibrium thermal structure shown in Figure~\ref{fig:climaTp}.  Flux profiles were computed 
       using the {\tt SMART} full-physics model (solid) and the radiative transfer routines adopted 
       in the {\tt Clima} model (dashed).}}
  \label{fig:clima_fluxes}
\end{figure}

{
We further explore the accuracy of our climate calculations via a comparison to a widely-adopted, 
one-dimensional radiative convective model---the {\tt Clima} tool developed by Kasting and 
collaborators \citep{kasting88,kastingetal1993,kopparapuetal2013,ramirezetal2014}.  As with our Venus 
simulations, we adopt (i) the {\tt SMART} model as our full-physics tool for computing requisite 
layer radiative properties and their Jacobians, (ii) the LiFE approach to adapting radiative 
fluxes to changes in atmospheric structure as the simulation timesteps to equilibrium, and 
(iii) a mixing-length approach to convective heat transport.  The case the we simulate is a 
planet with a 1~bar pure carbon dioxide atmosphere orbiting at 1~au from the Sun.  The world 
has an identical radius to Earth, a surface gravity of 10~m~s$^{-2}$, and a gray surface albedo 
of 0.20.  Both models compute a planetary average solar heating rate using eight solar zenith 
angles.  The LiFE-based approach began with a 250~K isothermal atmospheric temperature profile, 
and used two calls to the full-physics model to determine the layer radiative properties and 
their temperature derivatives.

The equilibrium thermal structures determined by {\tt Clima} and our LiFE-based approach are 
shown in Figure~\ref{fig:climaTp}.  Both models find a surface temperature of 310~K.  The 
thermal structure profiles are also in agreement, although the {\tt Clima} model finds additional 
structure in the radiative portion of the profile as well as an upper stratosphere that is warmer by 
about 10~K.  For the equilibrium thermal structure determined by the {\tt Clima} model, 
Figure~\ref{fig:clima_fluxes} shows the net solar and thermal flux profiles as computed by 
the {\tt Clima} and {\tt SMART} models, whose core radiative transfer routines were last 
inter-compared by Kopparapu~et~al. \citep{kopparapuetal2013}.  The net solar flux profiles 
from these two models are in good agreement.  The net thermal flux profiles, however, show 
some more substantial differences, including a 6~W~m$^{-2}$ discrepancy at the top of the 
atmosphere.  The net flux profile from the {\tt SMART} full-physics model yields strong cooling 
above 300~Pa, and, in general, heating between 300~Pa and the top of the convective zone.  
These heating/cooling trends would work to bring the {\tt Clima}-derived thermal structure into 
closer agreement with the LiFE-derived temperature-pressure profile.

A key characteristic of the {\tt Clima} model is its relatively short runtime---equilibrium thermal 
structures can typically be determined in minutes, or within an hour for more complicated 
scenarios.  This can be compared to runtimes for the LiFE approach, where computation 
of layer radiative properties and their Jacobians using a full-physics model takes hours to 
tens of hours (depending on complexity), and the process of timestepping towards an 
equilibrium solution can take several hours (although a single call to our two-stream 
flux adding routines only takes seconds).  An asset of the LiFE approach is versatility, 
however, as a huge variety of gases and/or aerosols can be straightforwardly incorporated 
into simulations (without needing to compute associated $k$-coefficients, for example), and 
runtimes only scale linearly (or better) with the number of added gases.  Furthermore, the 
LiFE approach benefits from having its radiative flux profiles being grounded in a full-physics 
radiative transfer tool.  Nevertheless, the LiFE approach would certainly benefit from updates 
focused on decreasing model runtime.  These updates could include implementing analytic 
Jacobians within the full-physics model used to compute layer radiative properties and their 
derivatives \citep{spurr&christi2007}, as well as an implementation of a root finding algorithm 
for determining equilibrium atmospheric states (as opposed to our current time-stepping 
algorithm).  
}

%%%
\section{Summary}
\label{sec:summary}
%%%

Planetary climate models require accurate radiative fluxes that can be easily and quickly 
updated in response to changes in the atmospheric and surface state.  We have described 
a technique --- the Linearized Flux Evolution (LiFE) approach --- that pairs a full-physics 
radiative transfer model with an efficient two-stream flux adding scheme to rapidly and 
accurately adapt radiative flux profiles to state variations.  The full-physics model is responsible 
for compiling the optical properties of the surface and atmosphere, solving for the angle- and 
level-dependent radiation field, determining the transmissivity, absorptivity, and source 
terms for each atmospheric layer (which, collectively, we call the layer radiative properties), 
and also computing Jacobians for these layer radiative properties with respect to any 
variable aspects of the atmospheric and surface state.  We believe this model to be the 
first of its kind, although recent work in the pure-absorption limit has used a line-by-line 
approach to computing radiative fluxes in proposed atmospheres for early Mars 
\citep{wordsworthetal2017}.

Using linear theory and the Jacobians, we update the computed layer radiative properties 
to small changes in the atmospheric and surface state.  While radiances are known to be 
a strongly non-linear function of the state, the use of quasi-monochromatic layer radiative 
properties helps to improve the linearity of the problem.  Additionally, for the atmospheric 
temperature component of the state vector, we use an analytic approach to remove the 
non-linear Planck-derived portion of layer thermal source terms.  Layer radiative 
properties that have been updated to reflect a change in the atmospheric and surface state 
can then be translated into new radiative flux profiles using the two-stream adding technique 
we describe.

By applying the LiFE approach to Mars, Earth, and Venus, we demonstrate its versatility.  
For Mars, we derive and show examples of the layer radiative properties and their 
temperature Jacobians.  Then, for Earth, we allow the thermal structure to evolve in time, 
demonstrating how LiFE can be used to timestep an atmospheric to pure radiative equilibrium.  
Finally, our application of LiFE to Venus shows how the approach can be used within a 
1-D radiative-convective model.  Using mixing length theory to compute the convective 
fluxes and LiFE to find the radiative fluxes, we determine an equilibrium thermal structure 
and for Venus, and net solar and thermal flux profiles, that strongly resemble observations.  
Given these successful applications, we hope that the LiFE approach will prove useful to 
many problems in planetary and Earth science.

%%%
\section*{Acknowledgements}
%%%
The approach to LiFE was originally created by DC, and was implemented, refined, and 
applied by both TR and DC.  TR gratefully acknowledges support from the National Aeronautics 
and Space Administration (NASA) through the Sagan Fellowship Program executed by the 
NASA Exoplanet Science Institute. The research by DC described in this paper was carried out at 
the Jet Propulsion Laboratory, California Institute of Technology, under a contract with NASA.  
Both TR and DC would like to acknowledge support from the NASA Astrobiology Institute's Virtual 
Planetary Laboratory, supported by NASA under Cooperative Agreement No. NNA13AA93A.  
The results reported herein benefitted from collaborations and/or information exchange 
within NASA's Nexus for Exoplanet System Science (NExSS) research coordination network 
sponsored by NASA's Science Mission Directorate.  Government sponsorship acknowledged. The 
authors would like to thank M. Marley for a friendly review of this work, E.~Schwieterman and 
R.~Kopparapu for facilitating comparisons with the {\tt Clima} model, and V.~Meadows for 
long-term support and dedication to this project.

%%% Appendix %%%
\appendix
%%%
\section*{Appendix}
\renewcommand{\thesection}{A\Alph{section}}
\renewcommand{\thesubsection}{\Roman{subsection}}
\renewcommand{\thesubsubsection}{\roman{subsubsection}}
%%%

%%%
\subsection{Deriving the Flux Adding Relations}
%%%

%%%
\subsubsection{Combining Two Homogenous Layers}
\label{sec:append:homogenous}
%%%

We can write a set of expressions, similar to Equations~\ref{eqn:adding_j_up} and 
\ref{eqn:adding_j_dn}, for layer $j+1$ as
\begin{equation}
F^{+}_{j} = r_{j+1}F^{-}_{j} + t_{j+1}F^{+}_{j+1} + s^{+}_{j+1} \ ,
\label{eqn:adding_j+1_up}
\end{equation}
\begin{equation}
F^{-}_{j+1} = t_{j+1}F^{-}_{j} + r_{j+1}F^{+}_{j+1} + s^{-}_{j+1} \ .
\label{eqn:adding_j+1_dn}
\end{equation}

Combining layers $j$ and $j+1$ yields an inhomogeneous layer, with emergent fluxes 
$F^{+}_{j-1}$ and $F^{-}_{j+1}$ and incident fluxes $F^{-}_{j-1}$ and $F^{+}_{j+1}$.  We 
can use Equations~\ref{eqn:adding_j_dn} and \ref{eqn:adding_j+1_up} to eliminate 
$F^{-}_{j}$ and $F^{+}_{j}$ from Equations~\ref{eqn:adding_j_up} and 
\ref{eqn:adding_j+1_dn}, which yields
\begin{equation}
\begin{array}{ll}
F^{+}_{j-1} = & \left(r_{j} + \frac{r_{j+1}t_{j}^2}{1 - r_{j}r_{j+1}}\right)F^{-}_{j-1} + 
              \frac{t_{j}t_{j+1}}{1 - r_{j}r_{j+1}}F^{+}_{j+1} \\ & + 
              \frac{t_{j}r_{j+1}}{1 - r_{j}r_{j+1}}s^{-}_{j} + 
              \frac{t_{j}}{1 - r_{j}r_{j+1}}s^{+}_{j+1} + s^{+}_{j} \ ,
\end{array}
\label{eqn:adding_inhom_up_long}
\end{equation}
\begin{equation}
\begin{array}{ll}
F^{-}_{j+1} = & \frac{t_{j}t_{j+1}}{1 - r_{j}r_{j+1}}F^{-}_{j-1} + 
              \left(r_{j+1} + \frac{r_{j}t_{j+1}^2}{1 - r_{j}r_{j+1}}\right)F^{+}_{j+1} \\ & + 
              \frac{t_{j+1}}{1 - r_{j}r_{j+1}}s^{-}_{j} + 
              \frac{r_{j}t_{j+1}}{1 - r_{j}r_{j+1}}s^{+}_{j+1} + s^{-}_{j+1} \ .
\end{array}
\label{eqn:adding_inhom_dn_long}
\end{equation}

Equations~\ref{eqn:adding_inhom_up_long} and \ref{eqn:adding_inhom_dn_long}, which 
encompass two homogenous layers, are in the form of the relations for a single layer 
(e.g., Equations~\ref{eqn:adding_j_up} and \ref{eqn:adding_j_dn}), and can be written as
\begin{equation}
F^{+}_{j-1} = R_{j,j+1}^{+}F^{-}_{j-1} + T_{j,j+1}^{+}F^{+}_{j+1} + S_{j,j+1}^{+} \ ,
\label{eqn:adding_inhom_up_short}
\end{equation}
\begin{equation}
F^{-}_{j+1} = T_{j,j+1}^{-}F^{-}_{j-1} + R_{j,j+1}^{-}F^{+}_{j+1} + S_{j,j+1}^{-} \ ,
\label{eqn:adding_inhom_dn_short}
\end{equation}
where we have defined the properties of an inhomogeneous layer encompassing layers 
$j$ and $j+1$ as
\begin{equation}
R_{j,j+1}^{+} = r_{j} + \frac{r_{j+1}t_{j}^2}{1 - r_{j}r_{j+1}} \ ,
\end{equation}
\begin{equation}
T_{j,j+1}^{+} = \frac{t_{j}t_{j+1}}{1 - r_{j}r_{j+1}} \ ,
\end{equation}
\begin{equation}
R_{j,j+1}^{-} = r_{j+1} + \frac{r_{j}t_{j+1}^2}{1 - r_{j}r_{j+1}} \ ,
\end{equation}
\begin{equation}
T_{j,j+1}^{-} = \frac{t_{j}t_{j+1}}{1 - r_{j}r_{j+1}} \ ,
\end{equation}
\begin{equation}
S_{j,j+1}^{+} = \frac{t_{j}r_{j+1}}{1 - r_{j}r_{j+1}}s^{-}_{j} + 
                \frac{t_{j}}{1 - r_{j}r_{j+1}}s^{+}_{j+1} + s^{+}_{j} \ ,
\end{equation}
\begin{equation}
S_{j,j+1}^{-} = \frac{t_{j+1}}{1 - r_{j}r_{j+1}}s^{-}_{j} + 
                \frac{r_{j}t_{j+1}}{1 - r_{j}r_{j+1}}s^{+}_{j+1} + s^{-}_{j+1} \ .
\end{equation}

Unlike a homogenous layer, the combined inhomogeneous layer does not reflect or transmit 
upwelling and downwelling fluxes symmetrically, as is indicated by the ``+'' and ``-'' superscripts 
on the inhomogeneous layer reflectivity and transmissivity.  The ability of a homogenous layer to reflect 
(or transmit) flux doesn't depend on whether it is illuminated from above or below.  For a 
inhomogeneous layer, though, it can, for example, be more effective at reflecting (or transmitting) 
flux that is incident from below than flux that is incident from above.

%%%
\subsubsection{Adding a Homogenous Layer to the Base of an Inhomogeneous Layer}
\label{sec:theory:addingdown}
%%%

In downward adding, we determine the radiative properties of successfully thicker inhomogeneous 
layers by recursively adding a homogenous layer to the base of an inhomogeneous layer.  For 
homogenous layer $j$, Equations~\ref{eqn:adding_j_up} and \ref{eqn:adding_j_dn} are still 
valid.  For the inhomogeneous layer extending from the top of the atmosphere ($j=0$) to layer 
$j-1$, we define the inhomogeneous layer reflectivity and source term for adding 
downward, $R^{-}_{0,j-1}$ and $S^{-}_{0,j-1}$, in Equation~\ref{eqn:inhomo_layer_down_j-1}.
Note that the contribution to the downwelling flux from the top-of-atmosphere boundary condition 
(i.e., $F^{-}_{0}$) can be included in the source term at the top of the atmosphere, 
$s^{-}_{0} = S^{-}_{0,0}$.

Inserting Equation~\ref{eqn:adding_j_up} into the right hand side of 
Equation~\ref{eqn:inhomo_layer_down_j-1} and simplifying yields
\begin{equation}
F^{-}_{j-1} = \left( t_{j}R^{-}_{0,j-1}F_{j}^{+} + s^{+}_{j}R^{-}_{0,j-1} + S^{-}_{0,j-1} \right)/\left(1 - r_{j}R^{-}_{0,j-1}\right) \ ,
\end{equation}
and inserting this into Equation~\ref{eqn:adding_j_dn} gives us
\begin{equation}
F^{-}_{j} = \left( r_{j} + \frac{t_{j}^{2}R^{-}_{0,j-1}}{1 - r_{j}R^{-}_{0,j-1}} \right)F_{j}^{+} + 
             s^{-}_{j} + \frac{t_{j}\left( S^{-}_{0,j-1} + s^{+}_{j}R^{-}_{0,j-1} \right)}{1 - r_{j}R^{-}_{0,j-1}} \ .
\end{equation}
Note that this is in a similar form to Equation~\ref{eqn:inhomo_layer_down_j-1}, and can be written as 
\begin{equation}
F^{-}_{j} = R^{-}_{0,j}F_{j}^{+} + S^{-}_{0,j} \ ,
\label{eqn:inhomo_layer_down_j}
\end{equation}
with $R^{-}_{0,j}$ and $S^{-}_{0,j}$ defined as in Equations~\ref{eqn:inhomo_refl_down} and 
\ref{eqn:inhomo_source_down}, respectively.

For an entire model atmosphere, downward layer adding begins with the top homogenous layer and a 
pair of boundary conditions (for $r_{0}$ and $s^{-}_{0}$), following the description in 
Section~I(\ref{sec:append:homogenous}).  Adding these two layers yields the quantities $R_{0,1}^{-}$ and 
$S_{0,1}^{-}$ for an inhomogeneous layer.  Equations~\ref{eqn:inhomo_refl_down} and 
\ref{eqn:inhomo_source_down} then provide a recursive set of relations that define the reflectivity 
and source terms of successfully thicker inhomogeneous layers for downward adding.

%%%
\subsubsection{Adding a Homogenous Layer to the Top of an Inhomogeneous Layer}
\label{sec:theory:addingup}
%%%

Upward adding proceeds in a similar fashion to downward adding, except that homogenous layers 
are added to the top of inhomogeneous layers.  Again, Equations~\ref{eqn:adding_j_up} and 
\ref{eqn:adding_j_dn} are still valid for homogenous layer $j$.  For the inhomogeneous layer extending 
from the base of the atmosphere ($j=N-1$) to the bottom of layer $j$, we have 
Equation~\ref{eqn:inhomo_layer_up}, which defines the inhomogeneous layer reflectivity and source term for 
adding upward ($R_{j+1,N}^{+}$ and $S_{j+1,N}^{+}$, respectively).  

Inserting Equation~\ref{eqn:adding_j_dn} into Equation~\ref{eqn:inhomo_layer_up} and simplifying 
yields
\begin{equation}
F_{j}^{+} = \left( t_{j}R_{j+1,N}^{+}F_{j-1}^{-} + s_{j}^{-}R_{j+1,N}^{+} + S_{j+1,N}^{+} \right)/\left( 1 - r_{j}R_{j+1,N} \right) \ .
\end{equation}
Inserting this equality for $F_{j}^{+}$ into Equation~\ref{eqn:adding_j_up} and simplifying then 
gives us
\begin{equation}
F_{j-1}^{+} = \left( r_{j} + \frac{t_{j}^{2}R^{+}_{j+1,N}}{1 - r_{j}R^{+}_{j+1,N}} \right)F_{j-1}^{-} + 
              s^{+}_{j} + \frac{t_{j}\left( S^{+}_{j+1,N} + s^{-}_{j}R^{+}_{j+1,N} \right)}{1 - r_{j}R^{+}_{j+1,N}} \ .
\end{equation}
This is in a similar form to Equation~\ref{eqn:inhomo_layer_up}, and can be written as 
\begin{equation}
F_{j-1}^{+} = R_{j,N}^{+}F_{j-1}^{-} + S_{j,N}^{+} \ ,
\end{equation}
with the recursive definitions for $R_{j,N}^{+}$ and $S_{j,N}^{+}$ given by 
Equations~\ref{eqn:inhomo_refl_up} and \ref{eqn:inhomo_source_up}, respectively.

Beginning with the bottom homogenous layer of the atmosphere and a pair of boundary conditions 
(for $r_{N}$ and $s_{N}^{+}$), we produce an inhomogeneous layer according to the 
description in Section~I(\ref{sec:append:homogenous}), thus yielding the quantities $R_{N-1,N}^{+}$ and 
$S_{N-1,N}^{+}$.  Equations~\ref{eqn:inhomo_refl_up} and \ref{eqn:inhomo_source_up} then provide 
a recursive set of relations that define the reflectivity and source terms of successfully thicker 
inhomogeneous layers for upward adding.

%%%
\subsection{Worked Example of Flux Adding and Linearized Evolution}
%%%

The flux adding scheme outlined in Section~\ref{sec:flxadding} is typically performed 
at a particular wavelength within a spectral interval.  However, insight 
can be gained by applying the flux adding method in a case where the 
optical properties of the atmosphere are gray (i.e., independent of 
wavelength).  For simplicity, we only consider thermal sources and we ignore scattering 
in this example.  { Also, we stress that the example below is designed to build 
intuition, and is not representative of the more complex approach described and used 
in this manuscript.}

We divide the atmosphere into $N-1$ layers, with boundaries at gray thermal optical depths 
given by $\tau_0$, $\tau_1$, ... $\tau_{N-1}$, with $\tau_{N-1} = \tau^{*}$, where $\tau^*$ 
is the total gray infrared optical depth of the atmosphere.  Note that we have 
included the diffusivity factor scaling \citep{rodgers&walshaw1966,armstrong1968} in our 
definition of the optical depth.  We take the flux transmissivity of layer $j$ to be given by
\begin{equation}
t_j = e^{- \Delta \tau_{j}} \ ,
\label{eqn:trns_ex}
\end{equation}
where $\Delta \tau_{j} = \tau_j - \tau_{j-1}$, and a simple expression of the layer source terms 
could be taken as 
\begin{equation}
s^{+,-}_j = \left( 1 - e^{-  \Delta \tau_{j}} \right) \sigma \bar{T}^4_{j} \ ,
\label{eqn:src_ex}
\end{equation}
where $\bar{T}_{j}$ is a representative average temperature of layer $j$, and the term 
in parentheses is the layer emissivity.  { Note that, here, we have assumed a functional 
form for the layer source terms, whereas in the manuscript above these are determined 
via comparisons to a full-physics model.}

By construction, the reflectivity of each layer is zero (i.e., $r_j = 0$), which greatly simplifies 
the adding approach, giving 
\begin{equation}
\begin{array}{lr}
R^+_{j,N}=R^-_{0,j}=0 & \forall j \ .
\end{array}
\label{eqn:Rzero}
\end{equation}
By inserting this into Equations~\ref{eqn:inhomo_layer_down_j} and \ref{eqn:upflx_adding}, we 
see that the upwelling and downwelling thermal fluxes (in this example) are simply given by the 
upward and downward adding source terms for the inhomogeneous layers, 
\begin{equation}
F^{-}_{j} = S^{-}_{0,j}
\end{equation}
\begin{equation}
F_{j}^{+} = S^{+}_{j+1,N}  \ .
\end{equation}
Additionally, the recursive relationships for the inhomogeneous layer source terms 
(Equations~\ref{eqn:inhomo_source_down} and \ref{eqn:inhomo_source_up}) simplify to 
\begin{equation}
S^{-}_{0,j} = s^{-}_{j} + t_{j}S^{-}_{0,j-1}
\label{eqn:inhomo_source_dnex}
\end{equation}
\begin{equation}
S_{j,N}^{+} = s^{+}_{j} +t_{j}S^{+}_{j+1,N} \ .
\label{eqn:inhomo_source_upex}
\end{equation}

When adding downwards, we start with the boundary condition that the downwelling thermal 
flux at the top of the atmosphere is zero,
\begin{equation}
F^{-}_{0} = S^{-}_{0,0} = 0 \ .
\end{equation}
We then use Equation~\ref{eqn:inhomo_source_dnex} to find:
\footnotesize
\begin{equation}
\begin{array}{ccccl}
S^{-}_{0,1} & = & s^{-}_{1} + t_{1}S^{-}_{0,0} & = & \left( 1 - e^{-  \Delta \tau_{1}} \right) \sigma \bar{T}^4_{1} \\ 
S^{-}_{0,2} & = & s^{-}_{2} + t_{2}S^{-}_{0,1} & = & \left( 1 - e^{-  \Delta \tau_{2}} \right) \sigma \bar{T}^4_{2} + 
                           e^{-  \Delta \tau_{2}}\left( 1 - e^{-  \Delta \tau_{1}} \right) \sigma \bar{T}^4_{1} \\
& & & \vdots & \\ 
S^{-}_{0,N-1} & = & s^{-}_{N-1} + t_{N-1}S^{-}_{0,N-2} & = & \left( 1 - e^{-  \Delta \tau_{N-1}} \right) \sigma \bar{T}^4_{N-1} + e^{-  \Delta \tau_{N-1}} \times \dots           
\end{array}
\end{equation}
\normalsize
Similarly, when adding upwards, we begin with the boundary condition that the upwelling flux at the base 
of the atmosphere is just the thermal flux from the surface, 
\begin{equation}
F_{N-1}^{+} = S^{+}_{N,N} = \sigma T_{s}^{4} \ ,
\end{equation}
and use Equation~\ref{eqn:inhomo_source_upex} to find:
\footnotesize
\begin{equation}
\begin{array}{ccccl}
S^{+}_{N-1,N} & = & s^{+}_{N-1} + t_{N-1}S^{+}_{N,N} & = & 
                            \ \left( 1 - e^{-  \Delta \tau_{N-1}} \right) \sigma \bar{T}^4_{N-1} + 
                                    e^{- \Delta \tau_{N-1}}\sigma T_{s}^{4} \\
S^{+}_{N-2,N} & = & s^{+}_{N-2} + t_{N-2}S^{+}_{N-1,N} & = & 
                            \ \left( 1 - e^{-  \Delta \tau_{N-2}} \right) \sigma \bar{T}^4_{N-2} + e^{- \Delta \tau_{N-2}} \times \\ & & & &
                              \left[  \left( 1 - e^{-  \Delta \tau_{N-1}} \right) \sigma \bar{T}^4_{N-1} + 
                              \sigma T_{s}^{4}e^{- \Delta \tau_{N-1}} \right] \\
& & & \vdots & \\
S^{+}_{1,N} & = & s^{+}_{1} + t_{1}S^{+}_{2,N} & = & \ \left( 1 - e^{-  \Delta \tau_{1}} \right) \sigma \bar{T}^4_{1} + e^{-  \Delta \tau_{1}} \times \dots
\end{array}
\end{equation}
\normalsize

The key element of the state vector in this example is the atmospheric temperature 
profile.  To determine how the flux profiles respond to changes in the temperature 
profile, we evaluate the derivatives of Equations~\ref{eqn:trns_ex} and 
\ref{eqn:src_ex} with respect to $\bar{T}_{j}$.  This gives us
\begin{equation}
\frac{\partial t_{j}}{\partial \bar{T}_{j}} = 0 \ ,
\end{equation}
\begin{equation}
\frac{\partial s^{+,-}}{\partial \bar{T}_{j}} = 4 \left( 1 - e^{-  \Delta \tau_{j}} \right) \sigma \bar{T}^3_{j} \ .
\end{equation}
In this example, the layer source term derivatives are analytic, and have a strong 
temperature dependence.

To proceed any further, we need to specify an atmospheric temperature profile, 
$T(\tau)$.  For simplicity, we adopt the radiative equilibrium temperature profile 
obtained by solving the two-stream Schwarzschild equation for the thermal 
radiative fluxes in a planetary atmosphere \citep[][p. 84]{andrews2010}, under 
the assumption that the atmosphere in transparent to shortwave radiation and 
that the thermal optical properties are gray:
\begin{equation}
T\left(\tau \right) = T_{skin} \left(1 + \tau \right)^{1/4} \ ,
\label{eqn:radeq_temp}
\end{equation}
where $T_{skin}$ is the skin temperature, which is given by
\begin{equation}
\sigma T_{skin}^4 = (1-A)F^{\odot}/8 \ ,
\end{equation}
where $A$ is the shortwave planetary Bond albedo, and $F^{\odot}$ is the 
top-of-atmosphere solar flux.  Additionally, the radiative equilibrium surface 
temperature is given by
\begin{equation}
T_{s} = T_{skin} \left(2 + \tau^* \right)^{1/4} \ ,
\end{equation}
and we simply take the average temperature for each model layer as 
\begin{equation}
\bar{T}_{j} = T \left( \frac{1}{2} (\tau_{j}+\tau_{j-1}) \right) \ .
\end{equation}

For this temperature profile, the layer source terms are given by
\begin{equation}
s^{+,-}_j = \left( 1 - e^{-  \Delta \tau_{j}} \right) 
    \left( 1 + \frac{1}{2}(\tau_{j}+\tau_{j-1}) \right) \sigma T_{skin}^4 \ ,
\label{eqn:ex_src}
\end{equation}
and their temperature Jacobians are
\begin{equation}
\frac{\partial s^{+,-}}{\partial \bar{T}_{j}} = 4 \left( 1 - e^{-  \Delta \tau_{j}} \right) 
     \left( 1 + \frac{1}{2}(\tau_{j}+\tau_{j-1}) \right)^{3/4} \sigma T_{skin}^3 \ .
\label{eqn:ex_srcderiv}
\end{equation}
If the temperature profile were different from the radiative equilibrium solution, 
with the temperature difference from this solution at each model layer given by 
$\delta \bar{T}_{j}$, then the thermal flux profiles would be determined by 
changing the layer source terms by an amount equal to
\begin{equation}
\delta s^{+,-} = \frac{\partial s^{+,-}}{\partial \bar{T}_{j}} \delta \bar{T}_{j} = 4 \left( 1 - e^{-  \Delta \tau_{j}} \right) 
     \left( 1 + \frac{1}{2}(\tau_{j}+\tau_{j-1}) \right)^{3/4} \sigma T_{skin}^3 \delta \bar{T}_{j} \ .
\end{equation}

\begin{figure}
  \centering
  \includegraphics[scale=0.6]{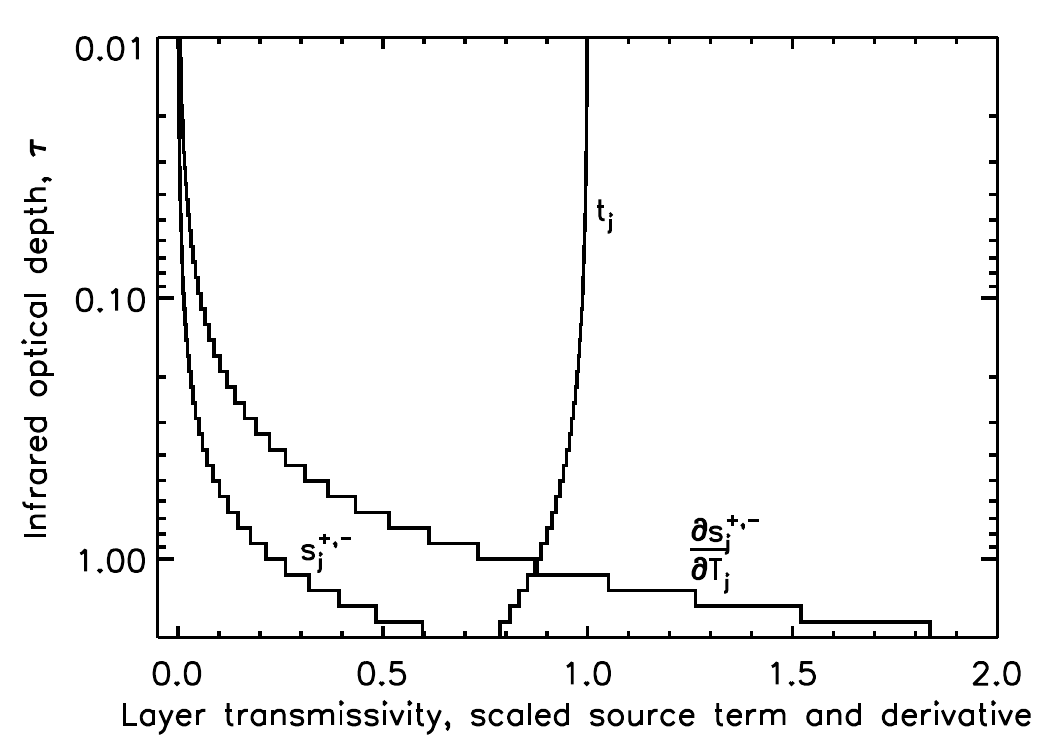}
  \includegraphics[scale=0.6]{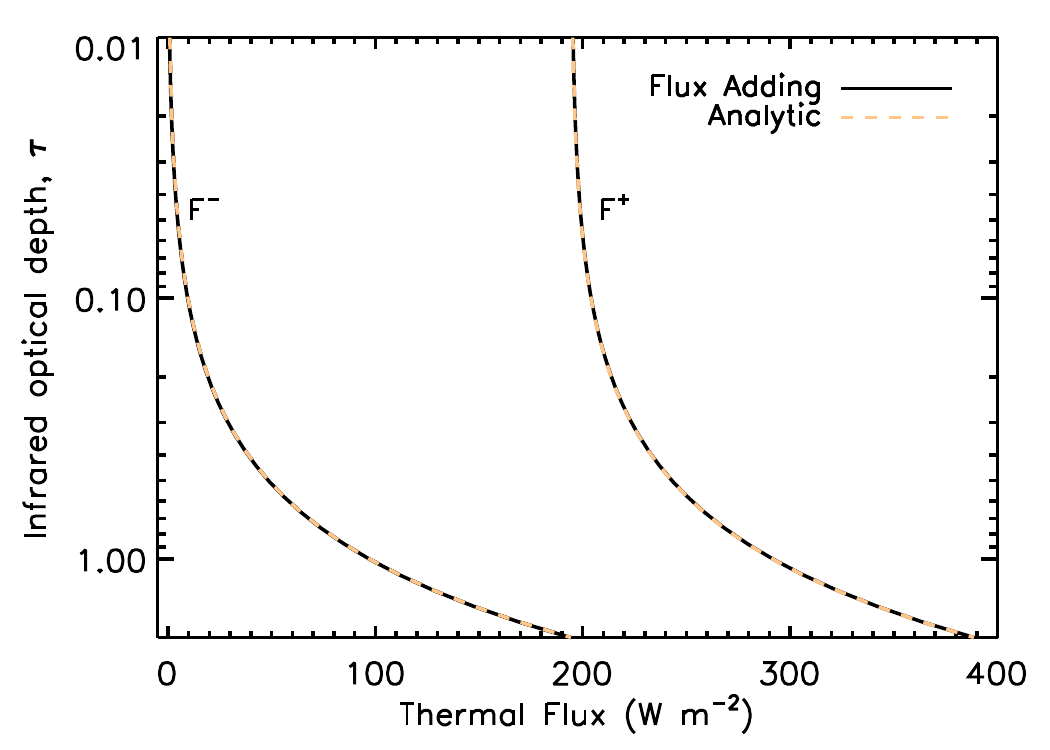}
  
  \caption{Layer radiative properties and thermal fluxes as a function of gray infrared optical depth 
                  from an example flux adding scenario.  The atmosphere is in radiative equilibrium, is 
                  transparent to solar radiation, and thermal radiation is treated according to the gray 
                  two-stream approximation.  This case has: $N=50$ atmospheric layers, a total 
                  atmospheric gray infrared optical depth $\tau^{*}=2$, and a skin temperature of 200~K.  
                  Layer transmissivities, $t_{j}$, and source terms, $s^{+,-}_{j}$, are from 
                  Equations~\ref{eqn:trns_ex} and \ref{eqn:ex_src}, respectively.  The source terms have been 
                  divided by $\sigma T_{skin}^{4}$.  The source term derivatives are from 
                  Equation~\ref{eqn:ex_srcderiv}, and have been scaled by $\sigma T_{skin}^{3}$.  The upwelling 
                  and downwelling fluxes, $F^{+}$ and $F^{-}$, are shown from the known analytic solution (dashed) 
                  and as computed from the flux adding approach (solid).}
  \label{fig:appendix:example}
\end{figure}

Figure~\ref{fig:appendix:example} demonstrates the results of this simple example for a 
particular case which has: 50 atmospheric layers, a total atmospheric gray infrared 
optical depth $\tau^{*}=2$, and a skin temperature of 200~K.  The atmospheric layers 
spaced in equal log units between $\tau_{0}=\tau^{*}/1000$ and the surface, which has a 
temperature of 288~K.  The layer transmissivities decrease towards the surface as a 
result of the logarithmic spacing (i.e., layers near the surface are more optically thick), and 
the layer source terms and their derivatives increase downwards because (1) the layers have 
higher emissivity, and (2) temperatures are increasing towards the surface.  The upwelling 
and downwelling thermal fluxes are in excellent agreement with the known analytic solution 
\citep[e.g.,][Equations (25) and (26)]{robinson&catling2012}.

%%%

%%%
%\bibliographystyle{elsarticle-num.bst}
%\bibliography{/Users/robinson/Documents/LaTeX/biblist}

%%%

%\clearpage

\newpage
\section{Tables and Figures}
\newpage

\begin{figure}
  \centering
  \includegraphics[scale=0.5]{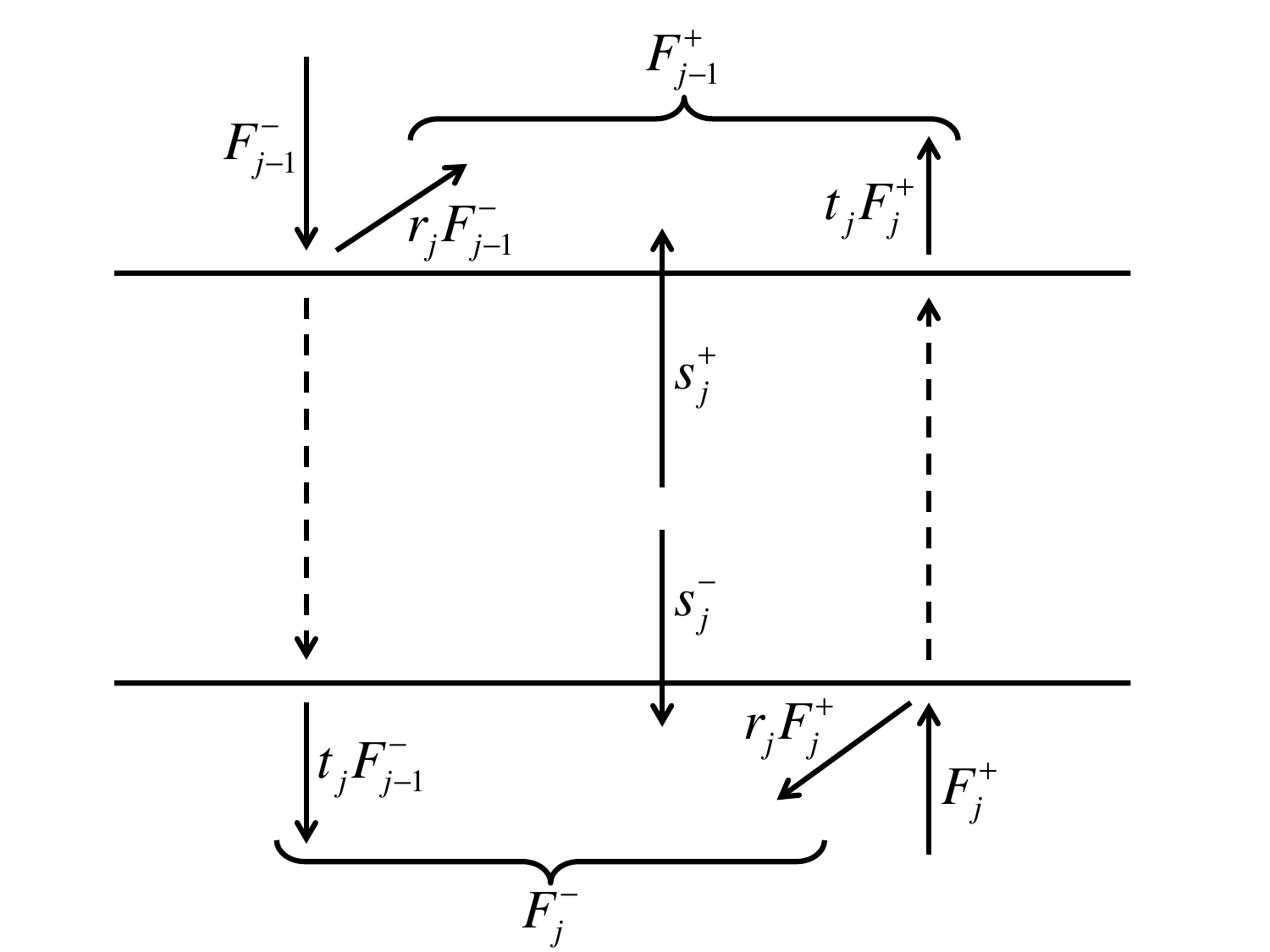}

  \caption{Schematic of layer radiative properties: the diffuse flux reflectivity ($r_{j}$), 
  transmissivity ($t_{j}$), and the layer source terms, $s^{+}_{j}$ and $s^{-}_{j}$.  The 
  fluxes ($F$) are defined at the layer boundaries, and are either upwelling ($+$) or 
  downwelling ($-$).  All terms are frequency-dependent.}
  \label{fig:adding_single}
\end{figure}

\begin{figure}
  \centering
  \includegraphics[scale=0.5]{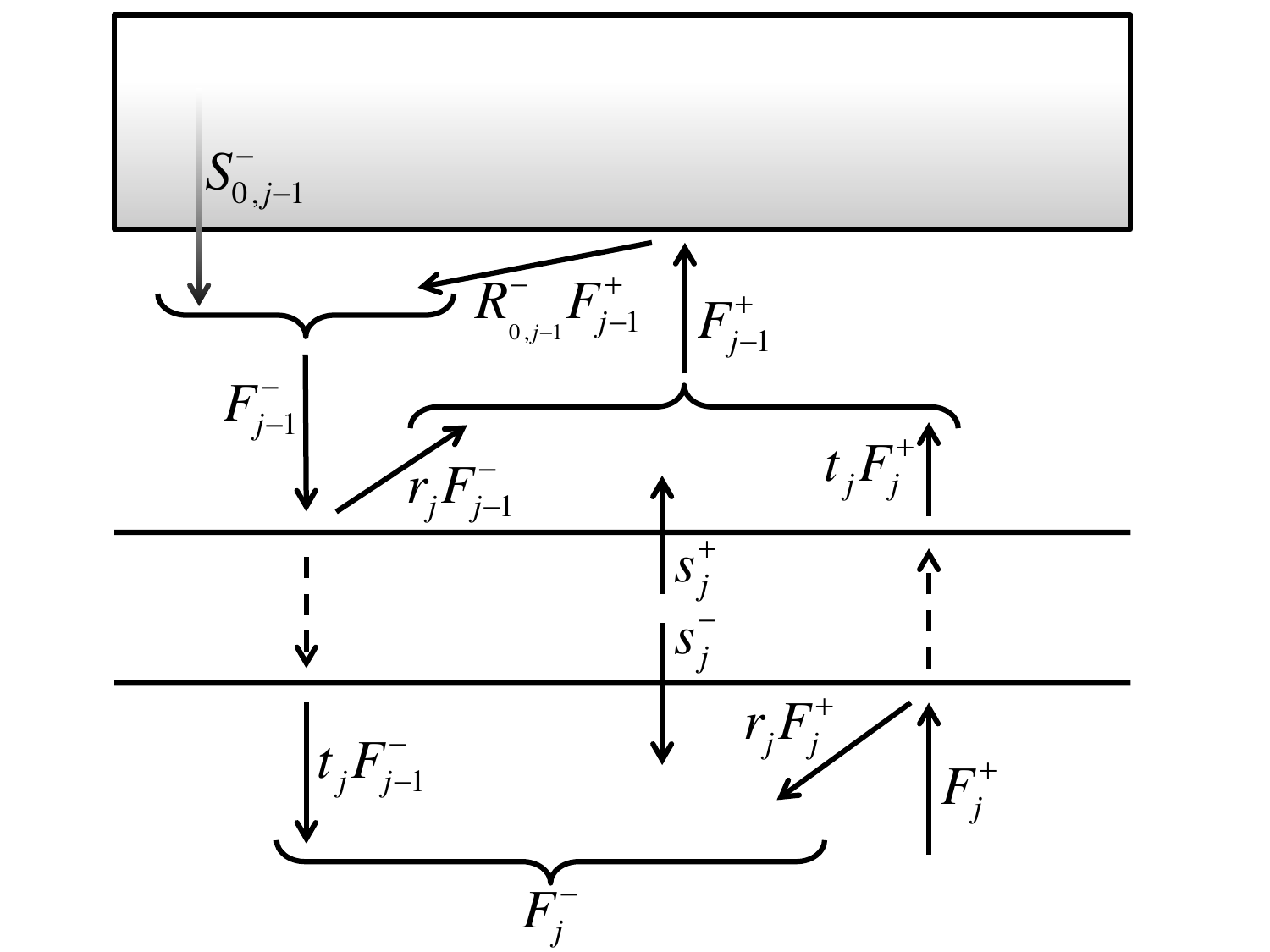}

  \caption{Schematic diagram of combining a homogenous and in homogenous layer by adding the 
           former to the top of the latter.  The inhomogeneous layer generates diffuse downwelling 
           flux through a source term, $S^{-}_{0,j-1}$, and reflects diffuse upwelling flux downward 
           through a reflectivity term, $R^{-}_{0,j-1}$.  Other symbols are as in 
           Figure~\ref{fig:adding_single}.}
  \label{fig:adding_down}
\end{figure}

\begin{figure}
  \centering
  \includegraphics[scale=0.5]{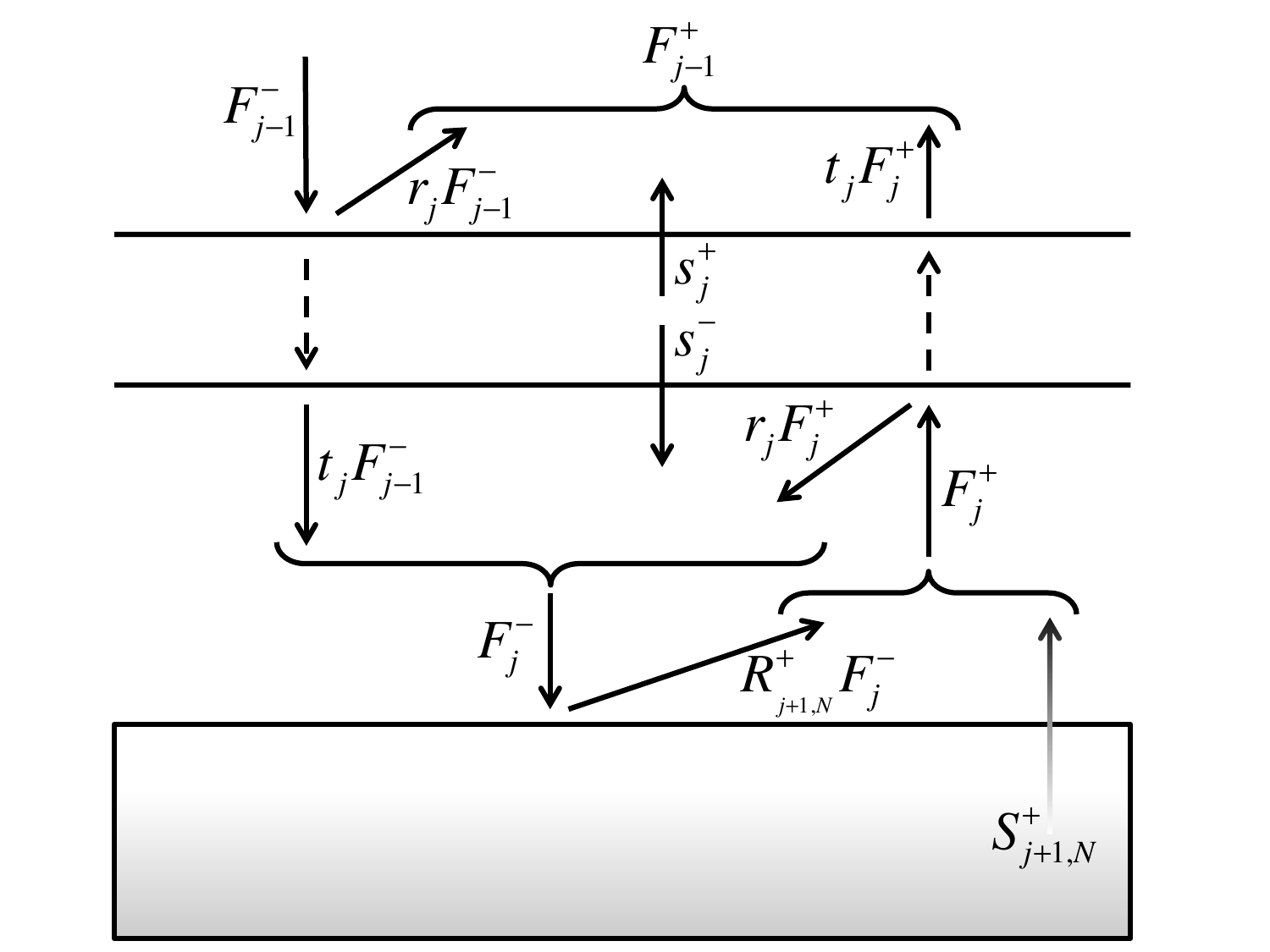}

  \caption{Schematic diagram of combining a homogenous and in homogenous layer by adding the 
           former to the bottom of the latter.  The inhomogeneous layer generates diffuse upwelling 
           flux through a source term, $S^{+}_{j+1,N}$, and reflects diffuse downwelling flux upward through a 
           reflectivity term, $R^{+}_{j+1,N}$.  Other symbols are as in 
           Figure~\ref{fig:adding_single}.}
  \label{fig:adding_up}
\end{figure}

%\begin{figure}
%  \centering
%  \includegraphics[scale=0.8]{figures/earth_rmix-eps-converted-to.pdf}
  
%  \caption{For Earth, standard planetary-average mass mixing ratio profiles for water vapor, carbon dioxide 
%  	and ozone \citep{mcclatcheyetal1972}.}
%  \label{fig:earth_rmix}
%\end{figure}

\clearpage

\clearpage

%\begin{figure}
%  \centering
%  \includegraphics[scale=0.8]{figures/demo_flux_diffs_frac-eps-converted-to.pdf}
%  
%  \caption{{ Fractional differences in net thermal flux profiles computed for perturbed temperature profiles 
%         (Figure~\ref{fig:demo_Tp}) when comparing our full-physics radiative transfer model and our 
%         Jacobian-based method.}}
%  \label{fig:demo_fracs}
%\end{figure}

%\begin{figure}
%  \centering
%  \includegraphics[scale=0.8]{figures/mars_temp-eps-converted-to.pdf}
%  
%  \caption{A standard Mars temperature-pressure profile \citep{conrathetal1973} used for computing 
%  	example layer radiative properties and their derivatives.}
%  \label{fig:mars_temp}
%\end{figure}

%\begin{figure}
%  \centering
%  \includegraphics[scale=0.8]{figures/mars_tau-eps-converted-to.pdf}
%  
%  \caption{A standard Mars dust extinction optical depth profile  \citep{conrath1975} used in our 
%  	calculations.}
%  \label{fig:mars_tau}
%\end{figure}

%\begin{figure}
%  \centering
%  \includegraphics[scale=0.8]{figures/mars_ssalb-eps-converted-to.pdf}
%  
%  \caption{Wavelength-dependent dust single-scattering albedo used in our calculations.}
%  \label{fig:mars_ssalb}
%\end{figure}

\begin{figure}
  \centering
  \includegraphics[scale=0.8]{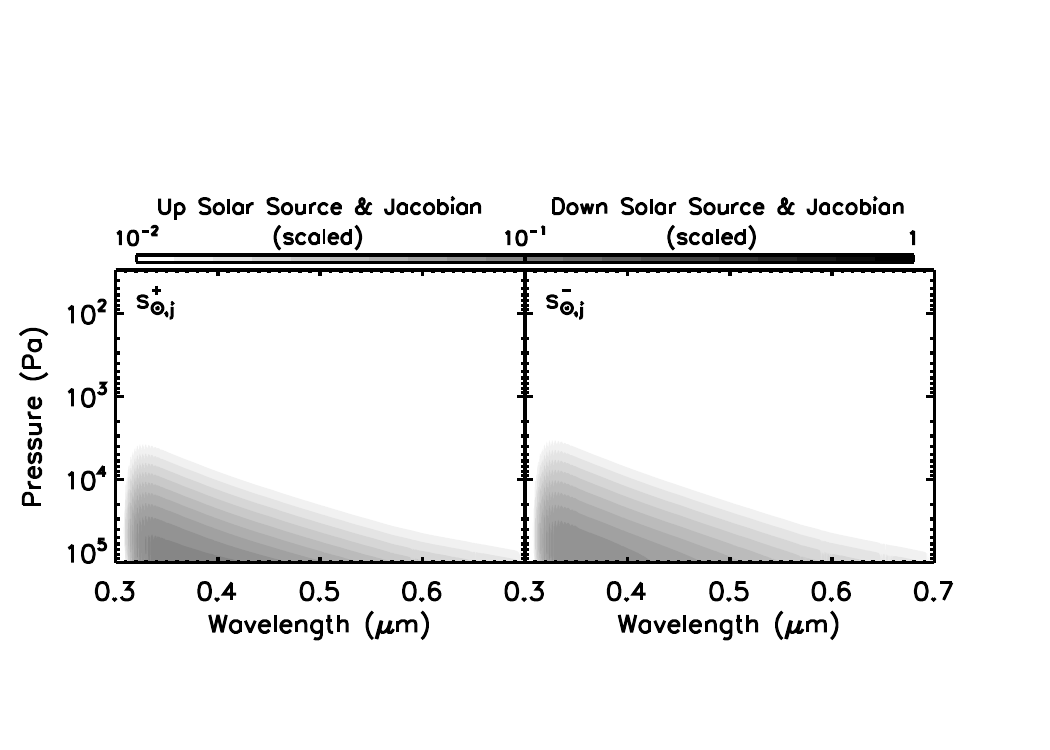}
  
  \caption{Initial wavelength-dependent upwelling (left) and downwelling (right) layer solar source terms (top) and 
  	their temperature derivatives (bottom) for Earth, used in our example application of the LiFE approach.  
	Both the source terms and their derivatives have been scaled by (1) the top-of-atmosphere solar flux, which 
	removes wavelength-dependent structure from the solar spectrum, and (2) $d\ln p_{j}$, which removes 
	layer path length effects and creates source terms that are smooth functions of pressure.  Shading is the 
	same as Figure~\ref{fig:mars_sol_source}.}
  \label{fig:earth_sol_source}
\end{figure}

\begin{figure}
  \centering
  \includegraphics[scale=0.8]{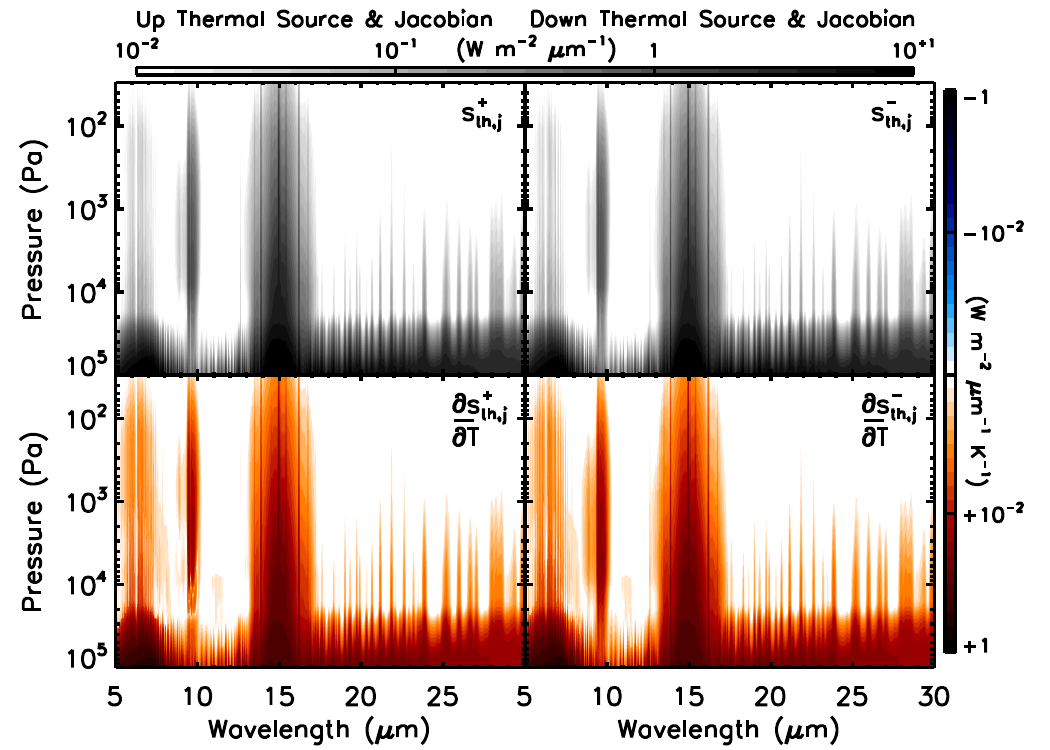}
  
  \caption{Initial wavelength-dependent upwelling (left) and downwelling (right) layer thermal source terms (top) and 
  	their temperature derivatives (bottom) for Earth, used in our example application of the LiFE approach.  Shading 
	is the same as Figure~\ref{fig:mars_therm_source}.}
  \label{fig:earth_therm_source}
\end{figure}

\end{document}